\documentstyle[12pt]{report}
\normalbaselines

\begin{document}
\bibliographystyle{unsrt}

\vbox {\vspace{16mm}} 
\begin{center}

{\LARGE \bf Radon transform and pattern functions    
\\[3mm] in quantum tomography}\\[7mm]                                         

Alfred W\"unsche\\ [2mm]
{\it Arbeitsgruppe "Nichtklassische Strahlung" der Max-Planck-Gesellschaft\\
Rudower Chaussee 5, 12489 Berlin, Germany}\\[3mm]
e-mail:wuensche@photon.fta-berlin.de\\
tel.: (4930) 6392 3982 \\
fax:  (4930) 6392 3990 \\[20mm]

\end{center}

\vspace{3mm}

\begin{abstract} 

   The two-dimensional Radon transform of the Wigner quasiprobability is 
introduced in canonical form and the functions playing a role in its inversion 
are discussed. The transformation properties of this Radon transform with 
respect to displacement and squeezing of states are studied and it is shown 
that the last is equivalent to a symplectic transformation of the variables 
of the Radon transform with the contragredient matrix to the transformation
of the variables in the Wigner quasiprobability. The reconstruction of the
density operator from the Radon transform and the direct reconstruction of 
its Fock-state matrix elements and of its normally ordered moments are 
discussed. It is found that for finite-order moments the integration over
the angle can be reduced to a finite sum over a discrete set of angles.
The reconstruction of the Fock-state matrix elements from the normally 
ordered moments leads to a new representation of the pattern functions by 
convergent series over even or odd Hermite polynomials which is appropriate 
for practical calculations. The structure of the pattern functions as first 
derivatives of the products of normalizable and nonnormalizable 
eigenfunctions to the number operator is considered from the point of view 
of this new representation.

\end{abstract}

\setcounter{chapter}{1}
\setcounter{equation}{0}
\chapter*{1. Introduction}
   
   The Radon transform of the Wigner quasiprobability is closely related 
to the expectation values or densities formed with the eigenstates to the 
rotated canonical observables. They possess 
all properties of genuine probability densities and can be measured since 
recent time in quantum optics of the radiation field by homodyne detection 
[1--4].
The field of problems of the 
reconstruction of the density operator from such or similar data is called 
quantum tomography. There arises the problem of the 
reconstruction of the density operator or of related quantities as matrix 
elements of the density operator or ordered moments from a complete set of 
tomographic data or of their approximate and, in some sense, optimal  
reconstruction from a more or less incomplete set of tomographic data.
The complete tomographic reconstruction of the density operator from the
Radon transform of the Wigner quasiprobability was prepared, as sometimes
happens with things in history, in a wondrous manner by a theoretical paper 
of Vogel and Risken \cite{vog} shortly before the experimental realisation 
of the corresponding measurements in quantum optics, although Vogel and 
Risken did not mention Radon transforms in this connection but speak about 
rotated quadrature phase. The transformation which is now called Radon 
transformation and which means the determination of field functions from
their integrated values over some manifolds was introduced by Radon in 1917 
but did not have a great impact on physics and also mathematics up to the 
fifties. Only, beginning from the sixties and prepared by mathematical papers
as \cite{vil} and the requirements of different tomographic problems the 
Radon transformation and its inversion became interesting for a greater 
community that is reflected in monographs \cite{helg} and review articles 
\cite{barr}. 

    The theoretical development in quantum tomography in the last years went
in the direction to determine more directly physical relevant parameters of 
the density operator as, for example, its matrix elements in the Fock-state 
basis 
[9--21]
or in the position representation \cite{welsch,khn} or 
to determine directly the normally ordered moments from tomographic data 
\cite{ri2,w7}. Other sets of measurable quantities as the displaced Fock-state 
matrix elements of the density operator with fixed displacement parameter
and varied excitation numbers are investigated for its appropriateness to
the reconstruction of the density operator \cite{opat}. Another direction is 
to study the influence of imperfect measurements onto the Radon transform 
of the Wigner quasiprobability. If this influence consists only in a smoothing
of the Wigner quasiprobability and of its Radon transform resulting in its 
convolution with a Gaussian function then there is, in principle, no loss of 
information and one has to take one of the smoothed quasiprobabilities 
instead of the Wigner quasiprobability. This was taken into account in some
of the first papers about reconstruction ( e.g., \cite{paul3} ) and was 
prepared already by the paper of Vogel and Risken \cite{vog} starting from
the rotated quadrature components of the $s$-orderd quasiprobabilities.
Another problem arises when there is a real loss or incomplete knowledge of
information. Then the reconstruction is no more unique and one needs a 
principle for an optimal reconstruction under such conditions. This problem
was recently studied from a general point of view in \cite{buzek} where the
incomplete knowledge was called observation level and this could become 
important in quantum tomography in the next time. The necessary principle is 
Jaynes' principle of maximal entropy and the best choice for the entropy in 
quantum mechanics is, in our opinion, for some formal reasons the 
Von-Neumann entropy.

The direct reconstruction of the matrix elements of the density operator 
in the Fock-state basis initiated by D'Ariano, Macchiavello and Paris 
\cite{ariano1} leads to the integration over the angle and over the 
position of the rotated quadrature components multiplied by specific 
functions of the angle and of the position coordinate. These auxiliary 
functions split into simple angle-dependent phase functions and more 
complicated position-dependent functions which were called pattern functions 
[12--14]
and were calculated in a representation by the 
functions of the parabolic cylinder. Then it was found that these pattern 
functions can be represented as first derivatives of products of the 
normalizable eigenfunction with one nonnormalizable eigenfunction of the 
number operator to eigenvalues corresponding to the numbers in the considered 
Fock-state matrix elements \cite{ri1}. The best way to recognize this special 
structure of the pattern functions is to consider the differential equations 
for the products of the eigenfunctions ( Hermite functions for the harmonic 
oscillator ) and to establish the corresponding orthogonality relations
by standard methods of the theory of ordinary differential equations that 
means with the help of the adjoint equations \cite{riw,kiss}.
	
The usual restriction of the representation of the Radon transform in 
quantum optics to the dependence on two variables ( rotation angle and 
line coordinate ) is unfavourable when considering the transformation 
properties with respect to squeezing of the states which results in symplectic
transformations of the arguments of the Wigner quasiprobability and of its
Radon transform. Therefore, in \cite{w7} was chosen a more general 
concept for the Radon transform as the starting point and it was explicitly
calculated this Radon transform for squeezed coherent states. The same more
general concept was taken in \cite{man1,man2} and called symplectic 
tomography and since the Radon transform contains complete information about
the quantum-mechanical state the basic equations of quantum mechanics were
reformulated for the Radon transform. This has the advantage that one works
directly with measurable quantities which can be considered as genuine
probability densities, at least, for each separate choice of the angle.

In the present paper we introduce in section 2 the two-dimensional Radon 
transform, as it seems to us, in its most rational form which we call
canonical form and consider its inversion where our intention is to clarify 
the meaning of some mathematical aspects which play a role in the inversion 
of two-dimensional Radon transforms and to give a new representations of this 
inversion. In section 3 we consider the transformation properties of the 
Wigner quasiprobability and of its Radon and Fourier transforms under 
displacement and squeezing transformations of initial states. The developed 
formulae, although highly technical and partially complicated, seem to us as 
very important for practical applications and theoretical calculations of 
the influence of squeezing that will be demonstrated for squeezed coherent 
states. Section 4 is devoted to the introduction and calculation of the 
pattern functions 
[12--14]
for the direct reconstruction of the 
Fock-state matrix elements from tomographic data and follows with weak 
deviations the main stream of considerations in the literature about quantum 
tomography but this section is also important as a preparation of the content 
of the next two sections. In section 5 we consider in a short form the direct 
reconstruction of the normally ordered moments from complete tomographic data 
without integration over the angle as recently published in our paper 
\cite{w7} and give some new formulae not represented there. Starting from the 
reconstruction of the density operator via the normally ordered moments we 
derive in section 6 an essentially new representation of the pattern
functions for the reconstruction of the Fock-state matrix elements of the 
density operator in form of series over Hermite polynomials of even or odd  
order which is an alternative for the practical calculation of
these functions to the existing representations by functions of the parabolic 
cylinder or by first derivatives of the product of the normalizable with a 
certain nonnormalizable eigenfunctions of the number operator in ``position''
representation. We find in this section a nonuniqueness of the pattern 
functions for the reconstruction of the Fock-state matrix elements. In 
section 7 we derive the specific structure of the pattern functions as
derivatives of a product of a normalizable and a nonnormalizable wave 
function from the fourth-order differential equation for products of Hermite
functions and its adjoint differential equation. Here we come again across
with the mentioned nonuniqueness of the pattern functions.

\setcounter{chapter}{2}
\setcounter{equation}{0}
\chapter*{2. Two-dimensional Radon and Fourier transforms and their inversion}

   The Wigner quasiprobability is the best compromise in quantum mechanics
for a phase-space description of a quantum-mechanical state in analogy to
the phase-space description in classical mechanics and statistics 
[30--38].
We consider one field mode corresponding to one mechanical degree of freedom 
and use a pair of canonical coordinates $(q,p)$ or complex coordinates 
$(\alpha,\alpha^*)$ and corresponding pairs of canonical operators $(Q,P)$ or 
boson annihilation and creation operator $(a,a^\dagger)$ in the following way
\begin{equation}
\alpha=\frac{q+ip}{\sqrt{2\hbar}},\quad \alpha^*=\frac{q-ip}{\sqrt{2\hbar}},
\quad a=\frac{Q+iP}{\sqrt{2\hbar}}, \quad a^\dagger=\frac{Q-iP}
{\sqrt{2\hbar}}, \quad \frac{i}{2} d\alpha \wedge d\alpha^* =\frac{dq \wedge dp}
{2\hbar}.
\end{equation}
In this section we consider the Radon transform of the Wigner quasiprobability
its connection to the Fourier transform of the Wigner quasiprobability and
the inversion of these transformations. All considerations of this section
are not specific for the Wigner quasiprobability and one can exchange the
symbol for the Wigner quasiprobability by the symbol for an arbitrary other
function of two variables over a plane, for example, another quasiprobability 
and one obtains in this way the general properties of two-dimensional Radon 
transformations. Specific properties of the Wigner quasiprobability and its 
Radon and Fourier transforms are considered from section 3 on. We use here 
the representation by the real variables $(q,p)$. All these relations can be 
easily translated into a representation by complex variables.

The Radon transform $\breve{W}(u,v;c)$ of the Wigner quasiprobability $W(q,p)$
can be defined in the following canonical form 
( compare [6--8] )
\begin{equation}
\breve{W}(u,v;c) \equiv \int dq \wedge dp \,\delta (c-uq-vp)W(q,p).
\end{equation}
The normalization of the Radon transform is closely connected to the 
normalization of the Wigner quasiprobability in the following way
\begin{equation}
\int_{-\infty}^{+\infty} dc\,\breve{W}(u,v;c)=\int dq \wedge dp \,W(q,p) = 1.
\end{equation}
Due to the relation
\begin{equation}
\breve{W}(\mu u,\mu v;\mu c) = \frac{1}{|\mu|} \breve{W}(u,v;c) , \qquad 
\mu \in R,
\end{equation}
where $\mu$ is an arbitrary real number this Radon transform depends 
effectively only on two continuous variables. The presence
of the delta function under the two-dimensional integral in (2.2) restricts
the integrations to the one-dimensional objects
\begin{equation}
uq+vp=c.    
\end{equation}
These are the equations for straight lines with $(u,v)$ as a normal vector
to the lines and $c/\sqrt{u^2+v^2}$ as a measure for the orthogonal 
( nearest ) distance of the line to the coordinate origin. However, 
this "oriented distance" can take on all real values contrary to positively 
definite distances. The coordinates $(u,v;c)$ are homogeneous line 
coordinates in the dual plane to the $(q,p)$-plane. 
In case of the generalization of the Radon transform to 
$N$-dimensional spaces the corresponding $(N-1)$-dimensional objects over 
which the field is integrated are the $(N-1)$-dimensional hyperplanes but 
this will be not considered here. Due to Eq.(2.2) the full information of 
the Radon transform is already contained in 
$\breve{W}(\cos\varphi,\sin\varphi;q_{\varphi})$ where 
$(u,v)=(\cos\varphi,\sin\varphi)$ is now the normal unit vector to the
straight lines. This reduced Radon transform is mostly denoted by
$w(\varphi,q_{\varphi})$ but we will see in the next section that such a
notation, although shorter, is unfavourable for the discussion of the 
transformation properties of the Radon transform. We call the Radon
transform as introduced in Eq.(2.2) the canonical form.

The Radon transform $\breve{W}(u,v;c)$ is closely related to the Fourier 
transform $\tilde{W}(u,v)$ of the Wigner quasiprobability $W(q,p)$ which can 
be defined by 
\begin{equation}
\tilde{W}(u,v) \equiv \int dq\wedge dp \, \exp\{-i(uq+vp)\}W(q,p), 
\end{equation}
with the inversion
\begin{equation}
W(q,p)=\frac{1}{(2\pi)^2}\int du\wedge dv \exp\{i(uq+vp)\}\tilde{W}(u,v),
\end{equation}
and with the normalization
\begin{equation}
\tilde{W}(0,0)=\int dq \wedge dp \,W(q,p)=1.
\end{equation}
The relation of the Radon transform to the Fourier transform is given by
\begin{equation}
\tilde{W}(u,v)=\int_{-\infty}^{+\infty} dc \,\exp(-ibc) \breve{W}\left(
\frac{u}{b},\frac{v}{b};c\right), \qquad b \in R,
\end{equation}
with arbitrary real numbers $b$ and its inversion is given by
\begin{equation}
\breve{W}(u,v;c)=\frac{1}{2\pi} \int_{-\infty}^{+\infty} db\; \exp(ibc)
\tilde{W}(bu,bv).
\end{equation}
The full inversion of the two-dimensional Radon transform can be made in two 
steps. The first step is the transition from the Radon transform to the 
Fourier transform and the second step is the inversion of the Fourier 
transform. If we introduce polar coordinates $(r,\varphi)$ instead of $(u,v)$ 
the following integral can be accomplished 
\begin{eqnarray}
\int_{-\infty}^{+\infty}dr\,|r|\,\exp\left(-ixr\right)\!\!&=&\!\!i\frac{
\partial}{\partial x}\left( \int_{0}^{+\infty}dr\,\exp\left(-ixr\right)-
\int_{-\infty}^{0} dr\,\exp\left(-ixr\right) \right) \nonumber\\ &=&\!\!
i\frac{\partial}{\partial x}\left(\frac{1}{i(x-i0)}+\frac{1}{i(x+i0)}\right) 
\nonumber\\ &=&\!\! 2 \frac{\partial}{\partial x}{\cal{P}}\frac{1}{x},
\end{eqnarray}
where ${\cal{P}}$ in front of $1/x$ means that integrals over the 
singularity of this function must be taken in the sense of Cauchy's principal 
value. Then we find
\begin{equation}
W(q,p)=\frac{r_0^2}{2\pi^2}\int_{0}^{\pi}d\varphi\,\int_{-\infty}^{+\infty} 
dc\,\left(\frac{\partial}{\partial c}{\cal{P}}\frac{1}{c-r_0(q\cos\varphi+
p\sin\varphi)}\right) \breve{W}(r_0\cos\varphi,r_0\sin\varphi;c)\nonumber\\,
\end{equation}
with arbitrary real fixed $r_0$. The integral over $c$ can be represented 
after partial integration as the Hilbert transform of the derivative of the 
Radon transform with respect to the variable $c$. 
A more symmetrical way of the representation 
of the inversion formula can be obtained if first the integration over $c$ is
made in the two-step inversion via the Fourier transform. One finds
\begin{equation}
W(q,p)=\frac{1}{2\pi^2}\int du \wedge dv\, \left({\cal{P}}\frac{1}{c-uq-vp}
\right)^{\!(1)} |c|\,\breve{W}(u,v;c), \quad c\in R,
\end{equation}
where the upper index $(1)$ denotes the first derivative and where $c$
is an arbitrary real number. 

It seems that the meaning of $(\partial/\partial x) {\cal{P}}(1/x)$ is not 
very clear in the literature about quantum tomography. In particular,
one cannot write $-{\cal{P}}\left(1/x^2\right)$ as the result of this
differentiation. Let us therefore discuss this more in detail. One has 
to consider $(\partial/\partial x){\cal{P}}(1/x)$ as a generalized function 
or linear functional and has to define all its derivatives as generalized 
derivatives in the sense of the theory of generalized functions. We introduce 
here the symbol ${\cal{R}}$ for the canonical regularization of a singular 
function ( Gel'fand and Shilov \cite{shilov} use the symbol $k.r.$ in the 
Russian and $CR$ in the English edition for canonical regularization ).
Canonical regularizations of different singular functions are synchronized
in the sense that with respect to additions of functions, multiplication of
functions by numbers and by well-behaved smooth functions and differentiations 
the regularization symbol can be changed in its order with these operations. 
The canonical regularization of the singular function $1/x$ is identical with 
the principal value ${\cal{P}}(1/x)$. If $\varphi(x)$ denotes arbitrary basis
functions then one can define ${\cal{R}}(1/x)$ by the following linear 
functional ( we do not specify in this physical context the space of basis 
functions which must be sufficiently smooth, in particular, at $x=0$ and 
sufficiently decreasing in infinity together with their derivatives )
\begin{eqnarray}
&&\left({\cal{R}}\frac{1}{x},\,\varphi(x) \right) \equiv 
\int_{-\infty}^{+\infty} dx \,\left({\cal{R}}\frac{1}{x}\right)\varphi(x) 
\equiv \int_{0}^{+\infty}dx \, \frac{\varphi(x)-\varphi(-x)}{x},\nonumber\\
&& \int_{-\infty}^{+\infty} dx \,\left({\cal{P}}\frac{1}{x}\right)\varphi(x)
\equiv \lim_{\varepsilon \to 0} \int_{|x|\ge\varepsilon} dx\,
\frac{\varphi(x)}{x},\qquad 
{\cal{R}}\frac{1}{x} \equiv {\cal{P}}\frac{1}{x} = \frac{\partial}{\partial x} 
\log|x|.
\end{eqnarray}
The function $\log |x|$ is a regular function because it is locally integrable
at $x=0$ and all canonical regularizations of power functions $1/x^n$
can be defined as derivatives of $\log|x|$ ( Gel'fand et al. \cite{vil,shilov} 
omit the regularization symbol ${\cal R}$ when discussing the functions 
${\cal R}(1/x^n)$ but understand these functions in the sense of canonical 
regularization ). The canonical regularization of $1/x^2$ can be defined by
\begin{eqnarray}
&&\left({\cal{R}}\frac{1}{x^2},\,\varphi(x) \right) \equiv 
\int_{-\infty}^{+\infty} dx \,\left({\cal{R}}\frac{1}{x^2}\right)\varphi(x) 
\equiv \int_{0}^{+\infty}dx \, \frac{\varphi(x)+\varphi(-x)-2\varphi(0)}{x^2},
\nonumber\\ &&{\cal{R}}\frac{1}{x^2} = -\frac{\partial}{\partial x} 
{\cal{R}}\frac{1}{x} = -\frac{\partial^2}{\partial x^2}\log|x|.
\end{eqnarray}
The proof of this formula can be given by the following chain of 
transformations including partial integration and using (2.14)
\begin{eqnarray}
\left({\cal{R}}\frac{1}{x^2},\,\varphi(x) \right)\!\!&=& \!\! 
-\left(-{\cal{R}}\frac{1}{x},\,\varphi^{(1)}(x) \right) \nonumber\\&=&\!\! 
\int_{0}^{+\infty} dx \,\frac{\varphi^{(1)}(x) - \varphi^{(1)}(-x)}{x} 
\nonumber\\&=&\!\! \int_{0}^{+\infty} dx\, \frac{1}{x}\frac{\partial}
{\partial x}\Big(\varphi(x)+\varphi(-x)-2\varphi(0)\Big)
\nonumber\\&=&\!\! -\int_{0}^{+\infty}dx \,\frac{\varphi(x)+\varphi(-x)
-2\varphi(0)}{-x^2} 
-\left\{\frac{\varphi(x)+\varphi(-x)-2\varphi(0)}{x}\right\}_{x=0}.\nonumber\\
\end{eqnarray}
The last expression in braces taken at $x=0$ vanishes and thus Eq.(2.15) is 
proved. All other proofs of canonical regularizations can be made in analogy
to this proof which is here given as an example for such kind of proofs.
The meaning of $\left({\cal{P}}(1/x)\right)^{(1)}$ in the inversion formulas 
for the Radon transform is therefore that we have to substitute it by 
$-{\cal{R}}(1/x^2)$ considered in the sense of Eq.(2.15). One may think 
that this does not give the right sign but it is not so. As an
example we consider $\varphi(x)=\exp(-x^2)$
\begin{eqnarray}
&&\left({\cal{R}}\frac{1}{x^2},\,\exp(-x^2) \right) \equiv
\int_{0}^{+\infty} dx\, \frac{2\big(\exp\left(-x^2\right)-1\big)}{x^2}
\nonumber\\ \!\!&=&\!\!
\int_{0}^{+\infty} dx\,\frac{-4x\exp\left(-x^2\right)}{x} \equiv
\left({\cal{R}}\frac{1}{x},\,-2x\exp(-x^2) \right) =
-2\sqrt{\pi}.
\end{eqnarray}
It could seem that ${\cal{R}}(1/x^2)$ and $\exp\left(-x^2\right)$ are both 
positively definite functions ( $\exp\left(-x^2\right)$ is really positively
definite ) but, nevertheless, the value of the linear functional in (2.17) 
for the considered basis function gives a negative number as shown. This 
means that ${\cal R}(1/x^2)$ is not a positively definite function.

The two-dimensional Radon transformation (2.2) together with its inversion 
(2.13) is equivalent to the following resolution of the two-dimensional 
identity operator 
\begin{equation}
-\frac{|c|}{2\pi^2}\int du \wedge dv \,\left({\cal R} \frac{1}{(c-uq-vp)^2}
\right)\,\delta(c-uq'-vp')=\delta(q-q')\delta(p-p').
\end{equation}
All the seeming difficulties with the inverse two-dimensional Radon 
transform can be avoided if the integration leading to the inversion
in two steps via the Fourier transform as the intermediate step is not 
made at an early stage of the calculations.

\setcounter{chapter}{3}
\setcounter{equation}{0}
\chapter*{3. Transformation properties of the Radon transform of the
Wigner quasiprobability} 

   The Wigner quasiprobability can be obtained by forming the trace over   
the product of the density operator with a transition operator. Among all  
quasiprobabilities it has the unique property that the operators for the 
transition from the density operator to the quasiprobability and back from 
the quasiprobability to the density operator via integration are the same.
The transition operators possess many equivalent representations. For our 
purposes we use the following convenient representation (e.g., \cite{w1} )
\begin{eqnarray}
&&W(q,p)=\left\langle \varrho 
\exp\left(-Q\frac{\partial}{\partial q}-P\frac{\partial}{\partial p} \right) 
\right\rangle \delta(q)\delta(p), \quad 
\int dq \wedge dp \,W(q,p)=1, \nonumber\\ && \varrho = 2\hbar \pi
\int dq \wedge dp \,W(q,p) 
\exp\left(-Q\frac{\partial}{\partial q}-P\frac{\partial}{\partial p} \right)  
\delta(q)\delta(p), \quad \langle A \rangle \equiv {\rm Trace}\{A\},  
\end{eqnarray}   
where we use the very rational notation $\langle A \rangle $ for the trace
of an operator $A$. We now investigate the transformation properties of the
Radon transform of the Wigner quasiprobability under displacement and
squeezing transformations of the density operator.

   Suppose that there is given a density operator $\varrho_{0}$ corresponding   
to a Wigner quasiprobability $W_{0}(q,p)$. Let us make a displacement of the   
state described by the density operator $\varrho_{0}$ by using the 
displacement operator $D(q,p)$ in the following way
\begin{equation}
\varrho = D(\bar{q},\bar{p})\varrho_{0}\big(D(\bar{q},\bar{p})\big)^\dagger,
\quad D(q,p)\equiv\exp\left\{-\frac{i}{\hbar}(qP-p\,Q)\right\}=
\big(D(-q,-p)\big)^\dagger.
\end{equation}   
Then we find for the Wigner quasiprobability $W(q,p)$ of the displaced 
state $\varrho$
\begin{eqnarray}
W(q,p)\!\!&=&\!\! \left\langle \varrho_{0}\big(D(\bar{q},\bar{p})
\big)^\dagger \exp\left(-Q\frac{\partial}{\partial q}-P\frac{\partial}
{\partial p} \right) D(\bar{q},\bar{p}) \right\rangle \delta(q)\delta(p)   
\nonumber\\&=&\!\! \left\langle \varrho_{0}
\exp\left\{-(Q+\bar{q}I)\frac{\partial}{\partial q} - (P+\bar{p}I) 
\frac{\partial}{\partial p}\right\} \right\rangle \delta(q)\delta(p) = 
W_{0}(q-\bar{q},p-\bar{p}),
\end{eqnarray}
meaning that the new Wigner quasiprobability is simply the 
displaced old Wigner quasiprobability in the phase plane. This property is 
also true for all other quasiprobabilities after displacement of the state 
due to the displacement structure of the quasiprobabilities themselves 
\cite{w1}. The Radon transform of the primary and of the transformed state 
are then connected by a displacement of the variable $c$
\begin{equation}
\breve{W}(u,v;c)=\breve{W}_{0}(u,v;c-u\bar{q}-v\bar{p}),
\end{equation}
whereas the Fourier transforms are connected by
\begin{equation}
\tilde{W}(u,v)=\exp\{-i(u\bar{q}+v\bar{p})\}\tilde{W}_{0}(u,v),
\end{equation}
that means by multiplication with a phase factor depending on $(u,v)$ and
on the displacement $(\bar{q},\bar{p})$.
   
   Our next aim is to consider the influence of unitary squeezing 
transformations of a state described by the density operator onto the Wigner 
quasiprobability and its Radon and Fourier transforms. We prepare this
by considering first some basic formulae for the, in general, nonunitary
squeezing operator $S(\xi,\eta,\zeta)$ which is defined by
\begin{eqnarray}
S(\xi,\eta,\zeta) \!\!&\equiv&\!\!\exp\left\{\frac{\xi}{2}a^2+i\frac{\eta}{2}
(a a^\dagger + a^\dagger a) -\frac{\zeta}{2}a^{\dagger\,2}\right\} \nonumber\\
&=&\!\! \exp\left\{\frac{i}{2\hbar}\left(\frac{-i(\xi-\zeta)+2\eta}{2}Q^2+
\frac{\xi+\zeta}{2}(QP+PQ)+\frac{i(\xi-\zeta)+2\eta}{2}P^2 \right)\right\}
\nonumber\\ &=&\!\! \big(S(-\zeta^*,-\eta^*,-\xi^*)\big)^\dagger. 
\end{eqnarray}
The transformation of the canonical operators $(Q,P)$ by the general  
squeezing operator is given by ( in representation by the boson operators 
$(a,a^\dagger)$ this is given in \cite{w2} and more specially in \cite{w4} )
\begin{eqnarray}
&&\!\!S(\xi,\eta,\zeta)(Q,P) \big(S(\xi,\eta,\zeta)\big)^{-1} \nonumber\\
&=&\!\!(Q,P)\left(\begin{array}{cc} 
{\rm ch} \varepsilon+{\displaystyle \frac{\xi+\zeta}{2}\, \frac{ {\rm sh}
\varepsilon}{\varepsilon}}, & 
{\displaystyle \frac{i(\xi-\zeta)-2\eta}{2}\,\frac{{\rm sh}\varepsilon}
{\varepsilon}}\\[2.5mm]
{\displaystyle \frac{i(\xi-\zeta)+2\eta}{2}\,\frac{{\rm sh}\varepsilon}
{\varepsilon}}, & 
{\rm ch} \varepsilon-{\displaystyle \frac{\xi+\zeta}{2}\, \frac{ {\rm sh}
\varepsilon}{\varepsilon}}  
\end{array}\right) \nonumber\\[2mm]
&\equiv&\!\! (Q,P)
\left(\begin{array}{cc} \alpha,& \beta\\ \gamma,& \delta \end{array} \right), 
\qquad
\left|\begin{array}{cc} \alpha,& \beta\\ \gamma,& \delta \end{array} \right|
=1, \qquad \varepsilon \equiv \sqrt{\xi \zeta -\eta^2},
\end{eqnarray}
with the inverse correspondence
\begin{eqnarray}
&&\!\!\xi=\frac{\alpha-i\beta-i\gamma -\delta}{2}\,
\frac{{\rm Arsh} \vartheta}{\vartheta}, \quad 
\eta=-\frac{\beta-\gamma}{2}\frac{{\rm Arsh} \vartheta}{\vartheta}, \quad
\zeta=\frac{\alpha+i\beta+i\gamma -\delta}{2}\,
\frac{{\rm Arsh} \vartheta}{\vartheta},\nonumber\\&& \vartheta \equiv
\sqrt{\left(\frac{\alpha+\delta}{2}\right)^2-1}=\pm {\rm sh}\varepsilon.
\end{eqnarray}
This is the fundamental two-dimensional representation of the $SL(2,C) \sim
Sp(2,C)$--group which in case of unitary squeezing operators $S(\zeta'^{\,*},
\eta'=\eta'^{\,*},\zeta')$ reduces two the $SL(2,R) \sim Sp(2,R) \sim 
SU(1,1)$--group \cite{pere,kim}. The matrices of this 
representation in (3.7) are complex unimodular or symplectic but in general 
not unitary. Symplectic transformations in spaces of even dimensionality 
are linear canonical transformations. They preserve an antisymmetric
2-form which can be nondegenerate only in spaces of even dimensionality 
whereas unimodular transformations preserve a fully antisymmetric $n$-form
( volume form ) in arbitrary $n$-dimensional spaces. Only in two-dimensional
spaces one has coincidence between symplectic and unimodular transformations.
The notions of symplectic transformations and symplectic groups were 
introduced by Weyl \cite{weyl}. We mention here that all problems concerning 
the multiplication of squeezing operators or the decomposition of squeezing 
operators into products with certain ordering are solvable by applying 
the fundamental representation of the two-dimensional symplectic or
unimodular group in Eqs.(3.7) and (3.8) ( see also \cite{mehta} and 
\cite{w1} ). 

In the special case 
\begin{equation}
\xi=\zeta=0, \quad \eta=\eta^*\equiv \varphi ,\quad S(0,\varphi,0)=\exp\left(
i\frac{\varphi}{2}\right) R(\varphi),\quad R(\varphi)\equiv \exp\left(
i\varphi a^\dagger a\right),
\end{equation}
this operation reduces to rotations according to 
\begin{equation}
R(\varphi)(Q,P)\big(R(\varphi)\big)^\dagger = (Q,P) \left( \begin{array}{cc}
\cos\varphi,& -\sin\varphi \\ \sin\varphi,& \cos \varphi \end{array} \right)
=(Q\cos\varphi + P\sin\varphi, -Q\sin\varphi + P\cos\varphi ), 
\end{equation}
where $R(\varphi)$ denotes the rotation operator. 

   We now consider the following squeezing transformation of the primary
state $\varrho_{0}$ by a unitary squeezing operator 
$S(\zeta'^{\,*},\eta'=\eta'^{\,*},\zeta')$ according to
\begin{equation}
\varrho=S(\zeta'^{\,*},\eta',\zeta')\varrho_{0}\big(S(\zeta'^{\,*},
\eta',\zeta')\big)^\dagger, \qquad  (\:\eta'=\eta'^{\,*}\:),
\end{equation}
where the unitary squeezing operator $S(\zeta'^{\,*},\eta'=\eta'^{\,*},
\zeta')$ is a special case of the, in general, nonunitary squeezing operator 
$S(\xi,\eta,\zeta)$ defined in Eq.(3.6). By using the transformation of the
basic canonical operators $(Q,P)$ given in (3.7) and taking into account
$\left(S(\zeta'^{\,*},\eta'=\eta'^{\,*},\zeta')\right)^\dagger =
S(-\zeta'^{\,*},-\eta'=-\eta'^{\,*},-\zeta')$ one finds
\begin{eqnarray}
W(q,p)\!\!&=&\!\! \left\langle \varrho_{0}\big(S(\zeta'^{\,*},\eta',
\zeta')\big)^\dagger \exp\left(-Q\frac{\partial}{\partial q}-
P\frac{\partial}{\partial p} \right) 
S(\zeta'^{\,*},\eta',\zeta') \right\rangle \delta(q)\delta(p)   
\nonumber\\&=&\!\! \left\langle \varrho_{0}
\exp\left\{-(Q\delta-P\gamma)\frac{\partial}{\partial q}-(-Q\beta+P\alpha) 
\frac{\partial}{\partial p}\right\} \right\rangle \delta(q)\delta(p) 
\nonumber\\&=&\!\! \left\{\left\langle \varrho_{0} 
\exp\left(-Q\frac{\partial}{\partial q'}-P\frac{\partial}{\partial p'}\right) 
\right\rangle \delta(q')\delta(p')\right\}_{(q'=\alpha q+\gamma p,\,p'=\beta q
+\delta p)}\nonumber\\[2mm]&=&\!\! W_{0}(\alpha q+\gamma p,\beta q+\delta p).
\end{eqnarray}
This means that the Wigner quasiprobability $W(q,p)$ for the state after
squeezing can be obtained from the initial Wigner quasiprobability 
$W_{0}(q,p)$ by an argument transformation with a real unimodular or 
symplectic matrix in the following way
\begin{eqnarray}
&& W(q,p)=W_{0}(\alpha q+\gamma p,\beta q+\delta p)\equiv W_{0}(q',p') ,\qquad
(q',p')=(q,p)\left(\begin{array}{cc} \alpha,& \beta\\ \gamma,& \delta 
\end{array} \right), \nonumber\\&&
\left(\begin{array}{cc} \alpha,& \beta\\ \gamma,& \delta \end{array} \right) 
=\left(\begin{array}{cc} 
{\rm ch} \varepsilon'+{\displaystyle \frac{\zeta'+\zeta'^{\,*}}{2}\, 
\frac{ {\rm sh}\varepsilon'}{\varepsilon'}}, & 
{\displaystyle \frac{-i(\zeta'-\zeta'^{\,*})-2\eta'}{2}\,\frac{{\rm sh}
\varepsilon'}{\varepsilon'}}\\[2.5mm]
{\displaystyle \frac{-i(\zeta'-\zeta'^{\,*})+2\eta'}{2}\,\frac{{\rm sh}
\varepsilon'}{\varepsilon'}}, & 
{\rm ch} \varepsilon'-{\displaystyle \frac{\zeta'+\zeta'^{\,*}}{2}\, 
\frac{ {\rm sh}\varepsilon'}{\varepsilon'}} \end{array}\right),\nonumber\\&&
\varepsilon'\equiv  \sqrt{\zeta' \zeta'^{\,*}-\eta'^{2}},\qquad
(\:\eta'=\eta'^{\,*}\:).
\end{eqnarray}
This result was originally obtained by Ekert and Knight \cite{ekert,knight}.

The Radon transform $\breve{W}(u,v;c)$ of the Wigner quasiprobability after 
squeezing of the state can be obtained from the Radon transform 
$\breve{W}_{0}(u,v;c)$ of the Wigner quasiprobability of the initial state 
by ( the double meaning of the symbol $\delta$ in the following formula 
can hardly lead to confusion )
\begin{eqnarray}
\breve{W}(u,v;c)\!\!&=&\!\! \int dq \wedge dp \,\delta(c-uq-vp)W(q,p)
\nonumber\\&=&\!\!\int dq \wedge dp \,\delta(c-uq-vp)W_{0}(\alpha q+\gamma p,
\beta q+\delta p) \nonumber\\&=&\!\!\int dq' \wedge dp' \,\delta\Big(
c-(\delta u- \beta v)q'-(-\gamma u+\alpha v)p'\Big) W_{0}(q',p')
\nonumber\\&=&\!\!\breve{W}_{0}\left(\delta u-\beta v,-\gamma u+\alpha v;
c\right),
\end{eqnarray}
which can be also written in the form 
\begin{eqnarray}  
&&\breve{W}(u,v;c) = \breve{W}_{0}\left(\delta u-\beta v,-\gamma u+ \alpha v
;c\right) \equiv \breve{W_{0}}(u',v';c),\nonumber\\[2mm] &&
\left(\begin{array}{c}u'\\v'\end{array}\right)=
\left(\begin{array}{rr} \delta,& -\beta\\ -\gamma,& \alpha \end{array}\right)
\left(\begin{array}{c}u\\v\end{array}\right)=
\left(\begin{array}{cc} \alpha,& \beta\\ \gamma,& \delta \end{array} 
\right)^{-1}\left(\begin{array}{c}u\\v\end{array}\right).
\end{eqnarray}
The transformation of the variables $(u,v)$ in the Radon transform is made
with the contragredient ( inverse transposed ) matrix to the matrix of the 
transformation of the canonical variables $(q,p)$ if we consider $(u,v)$ as
a row vector and by the inverse matrix if we consider $(u,v)$ as a column
vector. The same is true for the transformation of the variables $(u,v)$ of 
the Fourier transform for which one now easily finds
\begin{equation}  
\tilde{W}(u,v) = \tilde{W}_{0}\left(\delta u-\beta v,-\gamma u+
\alpha v\right) \equiv \tilde{W_{0}}(u',v').
\end{equation}
It could appear that it is more simple to derive first this result for
the Fourier transform of the Wigner quasiprobability and then the 
corresponding result for the Wigner quasiprobability itself but, really, both
derivations are of the same degree of difficulty and very similar.

The combination $uq+vp$ is invariant with respect to the considered 
unitary squeezing transformations of states 
\begin{equation}
u'q'+v'p'= (q',p')\left(\begin{array}{c}u'\\v'\end{array}\right) 
=(q,p)\left(\begin{array}{c}u\\v\end{array}\right) =uq+vp,
\end{equation}
but not with respect to displacements of states.
One can make both displacement and squeezing transformations of the initial
state but then the result for the transformed Wigner quasiprobability and
its Radon and Fourier transforms depends on the order of these operations.
If we make first the squeezing operation and after this the displacement then
we can immediately apply the derived transformation properties. However,
if we make first the displacement and after this the squeezing of the state
then we have to take into account that the displacement parameters must be
transformed by the real unimodular or symplectic matrices in the same way
as the corresponding coordinates. We do not explicitly write this down.
Usually, the Radon transform $\breve{W}(u,v;c)$ of the Wigner 
quasiprobability $W(q,p)$ is only considered in the special form $\breve{W}
(\cos\varphi,\sin\varphi;q_{\varphi})$ and denoted by 
$w(\varphi,q_{\varphi})$ but it is difficult to express such a fundamental 
transformation property as in Eq.(3.15) in the last notation. Let us also
mention that the transformation properties of all other quasiprobabilities 
and their Radon and Fourier transforms with respect to the considered 
squeezing transformations of states are not so simple as for the Wigner 
quasiprobability and we will not consider this here. 

Mancini, Man'ko and Tombesi \cite{man1,man2} consider practically the same 
object $\breve{W}(u,v;c)$ which is the Radon transform of the Wigner 
quasiprobability ( slightly extended by a further displacement ) and call
this symplectic tomography. Since $\breve{W}(u,v;c)$ contains the complete
information about the state one can transform the equations of time evolution 
for the state from any other quantity with complete information, for example, 
for the density operator or for the Wigner quasiprobability to an equation
for the Radon transform. A certain advantage of this approach is that we 
consider then directly the measurable quantities which are positively 
definite ( or semidefinite ) and can be considered for every fixed $(u,v)$
as genuine probability densities \cite{man1,man2}. The noncommutativity of 
the basic observables in quantum mechanics appears then as the, in general, 
noncompatibility of all these probability densities for essentially different 
$(u,v)$ on a classical level that means as marginals or Radon transforms of 
purely classical distribution functions $f(q,p)$. 

   Let us shortly consider the transformation of the Wigner quasiprobability
in the complex representation with respect to unitary squeezing 
transformations. The corresponding formula to (3.7) is then 
\begin{eqnarray}
&&\!\!S(\xi,\eta,\zeta)(a,a^\dagger) \big(S(\xi,\eta,\zeta)\big)^{-1} 
\nonumber\\&=&\!\!(a,a^\dagger)\left(\begin{array}{cc} 
{\rm ch} \varepsilon-{\displaystyle i\eta\, \frac{ {\rm sh}
\varepsilon}{\varepsilon}}, & {\displaystyle \xi\,\frac{{\rm sh}\varepsilon}
{\varepsilon}}\\[2.5mm]  {\displaystyle \zeta\,\frac{{\rm sh}\varepsilon}
{\varepsilon}}, & {\rm ch} \varepsilon+{\displaystyle i\eta\, 
\frac{ {\rm sh}\varepsilon}{\varepsilon}}  
\end{array}\right) \nonumber\\[2mm]
&\equiv&\!\! (a,a^\dagger)
\left(\begin{array}{cc} \kappa,& \lambda\\ \mu,& \nu \end{array} \right), 
\qquad
\left|\begin{array}{cc} \kappa,& \lambda\\ \mu,& \nu \end{array} \right|
=1, \qquad \varepsilon \equiv \sqrt{\xi \zeta -\eta^2},
\end{eqnarray}
with the correspondences
\begin{eqnarray}
&& \left(\begin{array}{cc}\kappa,& \lambda\\ \mu,& \nu \end{array} \right)
= \left(\begin{array}{cc}
{\displaystyle \frac{\alpha+i\beta-i\gamma+\delta}{2}},& 
{\displaystyle \frac{\alpha-i\beta-i\gamma-\delta}{2}}\\[2mm]
{\displaystyle \frac{\alpha+i\beta+i\gamma-\delta}{2}},&
{\displaystyle \frac{\alpha-i\beta+i\gamma+\delta}{2}}
\end{array}\right), \qquad \kappa+\nu = \alpha+\delta, \nonumber\\
&& \left(\begin{array}{cc}\alpha,& \beta\\ \gamma,& \delta \end{array} \right)
= \left(\begin{array}{cc}
{\displaystyle \frac{\kappa+\lambda+\mu+\nu}{2}},& 
{\displaystyle -i\frac{\kappa-\lambda+\mu-\nu}{2}}\\ [2mm]
{\displaystyle i\frac{\kappa+\lambda-\mu-\nu}{2}},&
{\displaystyle \frac{\kappa-\lambda-\mu+\nu}{2}}
\end{array}\right).
\end{eqnarray}
Then one finds the transformed Wigner quasiprobability after the unitary
squeezing transformation of the state according to (3.11) in analogy to (3.12)
\begin{equation}
W(\alpha,\alpha^*)= W_{0}(\nu^*\alpha+\mu \alpha^*,\mu^*\alpha + \nu \alpha^*)
\equiv W_{0}(\alpha',\alpha'^{\,*}),
\end{equation}
with the following specified unimodular or symplectic matrix due to the
unitarity of the considered state transformation
\begin{equation}
\left(\begin{array}{cc}\nu^*,& \mu^*\\ \mu,& \nu \end{array} \right)
=\left(\begin{array}{cc} 
{\rm ch} \varepsilon'-{\displaystyle i\eta'\, \frac{ {\rm sh}
\varepsilon'}{\varepsilon'}}, & {\displaystyle \zeta'^{\,*}\,\frac{{\rm sh}
\varepsilon'}
{\varepsilon'}}\\[2.5mm]  {\displaystyle \zeta'\,\frac{{\rm sh}\varepsilon'}
{\varepsilon'}}, & {\rm ch} \varepsilon'+{\displaystyle i\eta'\, 
\frac{ {\rm sh}\varepsilon'}{\varepsilon'}}  \end{array}\right), \quad
\varepsilon'\equiv \sqrt{\zeta'\zeta'^{\,*}-\eta'^{\,2}},\quad
( \eta'=\eta'^{\,*} ).
\end{equation}
The propagation of the transformation properties (3.20) of the Wigner 
quasiprobability in complex representation to its Radon and Fourier 
transforms in complex representation is easily to find if one introduces
a certain convention for the definition of these complex representations.
We will not do this here.

Let us now consider the more special unitary squeezing operator 
$S(\zeta'^{\,*},0,\zeta')$ as it is mostly used in the theory of squeezing
[48--58].
According to (3.6) we have
\begin{eqnarray}
S(\zeta'^{\,*},0,\zeta') \!\!&=&\!\! \exp\left\{ \frac{\zeta'^{\,*}}{2}\,
a^2 -\frac{\zeta'}{2}a^{\dagger\,2}\right\}\nonumber\\&=&\!\! 
\exp\left(-\frac{\zeta'}{2}\,\frac{{\rm th}|\zeta'|}{|\zeta'|}
a^{\dagger\,2}\right) \left({\rm ch}|\zeta'|\right)^{-(N+\frac{1}{2}I)}
\exp\left(\frac{\zeta'^{\,*}}{2}\,\frac{{\rm th}|\zeta'|}{|\zeta'|}a^{2}
\right) \nonumber\\&\equiv&\!\! 
\exp\left(-\frac{\zeta}{2}\,a^{\dagger\,2}\right)\left(\sqrt{1-|\zeta|^2}
\right)^{N+\frac{1}{2}I} \exp\left(\frac{\zeta^{\,*}}{2}\,a^{2}\right) ,
\quad N\equiv a^\dagger a,
\end{eqnarray}
where there is additionally given an almost normally ordered decomposition 
into factors ( the normal ordering of the central factor is mostly not 
necessary in applications but could be easily made ). Furthermore, it is made 
a new parametrization of the squeezing by the complex variable $\zeta$ 
characterized by 
\begin{equation}
\zeta=\zeta' \frac{{\rm th}|\zeta'|}{|\zeta'|},\quad 
\zeta'=\zeta \frac{{\rm Arth}|\zeta|}{|\zeta|},\quad
|\zeta|={\rm th}|\zeta'|,\quad |\zeta'|={\rm Arth}|\zeta|.
\end{equation}
The complex parameter $\zeta$ describes the squeezing operation in the
nonunitary approach \cite{w4,roy}
For convenience of applications of the preceding formulae we give here 
the explicit form of the corresponding specified matrices
\begin{eqnarray}
\left(\begin{array}{cc} \alpha,& \beta\\ \gamma,& \delta \end{array} \right) 
\!\!&=&\!\!\left(\begin{array}{cc} 
{\rm ch} |\zeta'|+{\displaystyle \frac{\zeta'+\zeta'^{\,*}}{2}\, 
\frac{ {\rm sh}|\zeta'|}{|\zeta'|}}, & 
{\displaystyle -i\frac{\zeta'-\zeta'^{\,*}}{2}\,\frac{{\rm sh}
|\zeta'|}{|\zeta'|}}\\[2.5mm]
{\displaystyle -i\frac{\zeta'-\zeta'^{\,*}}{2}\,\frac{{\rm sh}
|\zeta'|}{|\zeta'|}}, & 
{\rm ch} |\zeta'|-{\displaystyle \frac{\zeta'+\zeta'^{\,*}}{2}\, 
\frac{ {\rm sh}|\zeta'|}{|\zeta'|}} \end{array}\right)\nonumber\\
&=&\!\!\frac{1}{\sqrt{1-|\zeta|^2}}\left(\begin{array}{cc}
\displaystyle{1+\frac{\zeta+\zeta^*}{2}},& 
\displaystyle{-i\frac{\zeta-\zeta^*}{2}}\\[2mm]
\displaystyle{-i\frac{\zeta-\zeta^*}{2}},&
\displaystyle{1-\frac{\zeta+\zeta^*}{2}}\end{array}\right),\nonumber\\
\left(\begin{array}{cc}\nu^*,& \mu^*\\ \mu,& \nu \end{array} \right)
&=&\!\! \left(\begin{array}{cc} 
{\rm ch} |\zeta'|, & {\displaystyle \zeta'^{\,*}\,\frac{{\rm sh}|\zeta'|}
{|\zeta'|}}\\[2.5mm]  {\displaystyle \zeta'\,\frac{{\rm sh}|\zeta'|}
{|\zeta'|}}, & {\rm ch} |\zeta'| \end{array}\right)=
\frac{1}{\sqrt{1-|\zeta|^2}}\left(\begin{array}{cc} 1,& \zeta^*\\
\zeta,& 1 \end{array}\right).
\end{eqnarray}
Now we can apply all derived transformation formulae for the Wigner 
quasiprobability and its Radon and Fourier transforms and have to specify the
corresponding matrix elements by the expressions given here. 

The substitution of the complex squeezing parameter $\zeta'$ by $\zeta$ 
according to Eq.(3.22) is not only a formal substitution but has a deep
physical content. If one acts with squeezing and displacement operators
onto pure states $|\psi\rangle$ one can represent $|\psi\rangle$ by its
Fock-state representation. Then one can represent the Fock states by the
action of powers of the creation operator onto the vacuum state $|0\rangle$
that means by an excitation operator. Then one can bring the squeezing 
operator of the form (3.22) by application of the commutation rules with
the excitation operator in the position that it acts in unchanged form 
directly onto the vacuum state. By using the ordered decomposition given
in (3.22) the operator $\exp\left((\zeta^*/2)a^2\right)$ acts onto the
vacuum state as the unity operator whereas 
$\left(\sqrt{1-|\zeta|^2}\,\right)^{N+(1/2)I}$ 
multiplies it with a normalization factor $\left(1-|\zeta|^2\right)^{1/4}$.
The essential action of the squeezing operator onto the vacuum state can 
be then substituted by the action of the nonunitary squeezing operator
$\exp\left(-(\zeta/2)a^{\dagger\,2}\right)$. After this the changed 
excitation and displacement operators act onto the obtained state that must 
be considered for the concrete case. This procedure should be also possible 
for mixed states or genuine density operators. Hence the squeezing parameter 
$\zeta$ in the nonunitary approch possesses a concrete physical meaning.
In case of the more general unitary squeezing operator $S(\zeta'^{\,*},
\eta'=\eta'^{\,*},\zeta')$ one has $1$ complex parameter $\zeta'$ and $1$ 
real parameter $\eta'$ that is more than the only $1$ complex parameter 
$\zeta$ in the nonunitary approach and there remains a certain freedom in 
the choice of the correspondence between unitary and nonunitary approaches
\cite{w2}. We do not discuss this here.                                             

As an example we consider the vacuum state $\varrho_{0}=|0\rangle \langle 0|$.
The Wigner quasiprobability and its Radon and Fourier transforms are 
\begin{eqnarray}
&&W_{0}(q,p)=\frac{1}{\hbar \pi}\exp\left\{-\frac{q^2+p^2}{\hbar}\right\}, 
\nonumber\\ && 
\breve{W}_{0}(u,v;c)=\frac{1}{\sqrt{\hbar\pi(u^2+v^2)}}\exp\left\{
-\frac{c^2} {\hbar(u^2+v^2)}\right\},\nonumber\\&& \tilde{W}_{0}(u,v)=
\exp\left\{-\frac{\hbar(u^2+v^2)}{4}\right\}.
\end{eqnarray}
After squeezing of the state with the special unitary squeezing operator 
in Eq.(3.23) and succeeding displacement one obtains squeezed coherent states
and by using Eqs.(3.3)--(3.5) and (3.13)--(3.16) as well as 
(3.24) one finds the following Wigner quasiprobability and its Radon and 
Fourier transforms \cite{w7}
\begin{eqnarray}
&&W(q,p)=\frac{1}{\hbar \pi}\exp\left\{-\frac{|1+\zeta|^2
(q-\bar{q})^2+|1-\zeta|^2(p-\bar{p})^2-i(\zeta-\zeta^*)2(q-\bar{q}) 
(p-\bar{p})}{\hbar\left(1-|\zeta|^2\right)}\right\}, \nonumber\\&&
\breve{W}(u,v;c)=\sqrt{\frac{1-|\zeta|^2}{\hbar\pi\left(|1-\zeta|^2
u^2+|1+\zeta|^2 v^2 +i(\zeta-\zeta^*)2uv\right)}}
\nonumber\\&& \qquad \qquad \qquad \!\! \times
\exp\left\{-\frac{
\left(1-|\zeta|^2\right)(c-u\bar{q}-v\bar{p})^2}{\hbar\left(|1-\zeta|^2 u^2+
|1+\zeta|^2 v^2+i(\zeta-\zeta^*)2uv\right)}\right\},\nonumber\\ &&
\tilde{W}(u,v)= \exp\left\{-i(u\bar{q}+v\bar{p})\right\}
\nonumber\\[2mm] && \qquad \qquad \quad \! \times
\exp\left\{-\frac{\hbar\left(|1-\zeta|^2 u^2+ |1+\zeta|^2 v^2
+i(\zeta-\zeta^*)2uv\right)}{4\left(1-|\zeta|^2\right)}
\right\}.
\end{eqnarray}
We mention the following complex factorizations of the numerators and  
denominators in the exponents which are possible with rational coefficients 
in $\zeta$ and $\zeta^*$ only for the Wigner quasiprobability and its 
Radon and Fourier transforms but not for other important quasiprobabilities 
of these states such as, for example, the coherent-state quasiprobability.
\begin{eqnarray}
&& |1+\zeta|^2 q^2 +|1-\zeta|^2 p^2-i(\zeta-\zeta^*) 2qp =
\big\{(1+\zeta)q+i(1-\zeta)p\big\}\big\{(1+\zeta^*)q-i(1-\zeta^*)p\big\},
\nonumber\\ 
&& |1-\zeta|^2 u^2 +|1+\zeta|^2 v^2+i(\zeta-\zeta^*) 2uv =
\big\{(1-\zeta)u+i(1+\zeta)v\big\}\big\{(1-\zeta^*)u-i(1+\zeta^*)v\big\}.
\nonumber\\
\end{eqnarray}
They express a certain duality of the transformation of the variables in
the Wigner quasiprobability on one side and its Radon and Fourier transform 
on the other side.

For convenience we give in the following formula the connection between our
parameters $\bar{q},\bar{p}$ and $\zeta$ for squeezed coherent states and
the expectation values of the canonical operators $(Q,P)$ and their 
dispersions and symmetrical correlation
( $\overline{A}\equiv \langle \varrho A \rangle,\; \Delta Q \equiv   
Q-\overline{Q}I,\; \Delta P\equiv P-\overline{P}I$ )
\begin{eqnarray}
&&\overline{Q} = \bar{q},\quad \overline{P} = \bar{p}, \quad
\overline{(\Delta Q)^2} =\frac{\hbar}{2} \frac{|1-\zeta|^2}
{1-|\zeta|^2},\quad 
\overline{(\Delta P)^2} =\frac{\hbar}{2} \frac{|1+\zeta|^2}
{1-|\zeta|^2}, \nonumber\\ && \frac{1}{2} \big(\,\overline{\Delta Q \Delta P}
+\overline{\Delta P \Delta Q}\,\big) = i\frac{\hbar}{2} \frac{\zeta-\zeta^*}
{1-|\zeta|^2}.
\end{eqnarray}
These 5 real parameters can be used for the unique characterization of 
squeezed coherent states. Because of
\begin{equation}
\overline{(\Delta Q(\varphi))^2}\;\overline{(\Delta P(\varphi))^2}
-\frac{1}{4}\Big(\,\overline{\Delta Q(\varphi) \Delta P(\varphi)}+
\overline{\Delta P(\varphi) \Delta Q(\varphi)}\,\Big)^2=
\frac{\hbar^2}{4},
\end{equation}
for arbitrary $\varphi$ they effectively reduce to 4 independent parameters 
for squeezed coherent states. However, for general Gaussian Wigner 
quasiprobabilities corresponding to displaced and squeezed thermal states
these parameters become independent and (3.39) has to be substituted by
an inequality corresponding to a modification of the usual uncertainty 
relations by inclusion of the uncertainty correlation 
[60-62].
The maximal and minimal values of $\sqrt{\overline{\left(\Delta Q(\varphi)
\right)^2}}$ for varying $\varphi$ are determined by
\begin{eqnarray}
&&\sqrt{\overline{(\Delta Q(\varphi_{max}))^2}} = \sqrt{\frac{\hbar}{2}\,
\frac{1+|\zeta|}{1-|\zeta|}},\quad
\sqrt{\overline{(\Delta Q(\varphi_{min}))^2}} = \sqrt{\frac{\hbar}{2}\,
\frac{1-|\zeta|}{1+|\zeta|}},\nonumber\\&&
\exp(i2\varphi_{max})=-\frac{\zeta}{|\zeta|},\quad 
\exp(i2\varphi_{min})=+\frac{\zeta}{|\zeta|},\; \longrightarrow \;
\exp(i4\varphi_{ext})=\frac{\zeta}{\zeta^*}.
\end{eqnarray}
These uncertainties can be considered as one of the possible definitions of 
the widths of the principal axes of the squeezing ellipse. 
For the corresponding angles $\varphi_{ext}$ determined by (3.30) the
symmetrical correlation vanishes, i.e.
\begin{equation}
\overline{\Delta Q(\varphi_{ext}) \Delta P(\varphi_{ext})}+
\overline{\Delta P(\varphi_{ext}) \Delta Q(\varphi_{ext})}=0.
\end{equation}

In an analogous manner one can consider squeezing and displacement of 
other initial states, for example, Fock states, thermal states or 
Schr\"{o}dinger cat states.

\setcounter{chapter}{4}
\setcounter{equation}{0}
\chapter*{4. Reconstruction of the density operator from the Radon transform}

In this section we consider the reconstruction of the density operator and    
of its matrix elements in the Fock-state basis from the Radon transform of the
Wigner quasiprobability. Let us first establish the following relation 
between the Radon transform and the rotated marginals of the Wigner 
quasiprobability
\begin{equation}   
\breve{W}(\cos\varphi,\sin\varphi;q)=\langle q|\big(R(\varphi)\big)^\dagger
\varrho R(\varphi) |q\rangle \equiv 
\langle q;\varphi|\varrho|q;\varphi\rangle,
\end{equation}
where $|q;\varphi \rangle$ denotes the orthonormalized eigenstates of the 
rotated canonical operator $Q(\varphi)$ to eigenvalues $q$ in the following 
way ( see Eq.(3.10) )
\begin{eqnarray}
&&Q(\varphi)|q;\varphi\rangle = R(\varphi)Q\big(R(\varphi)\big)^\dagger 
R(\varphi)|q\rangle = q|q;\varphi\rangle, \quad Q|q\rangle = q|q\rangle,
\quad |q;\varphi\rangle \equiv R(\varphi)|q\rangle, 
\nonumber\\&& \langle q;\varphi|q';\varphi \rangle = \delta(q-q'), \qquad 
\int_{-\infty}^{+\infty} dq\,|q;\varphi\rangle \langle q;\varphi| = I.
\end{eqnarray}
To prove Eq.(4.1) we first immediately find the special result                                           
\begin{equation}   
\langle q|\varrho|q \rangle = \int_{-\infty}^{+\infty}dp\,W(q,p) = \breve{W}
(1,0;q).
\end{equation}   
Then, we see from Eq.(4.1) that $\langle q;\varphi|\varrho|q;\varphi\rangle$
corresponds to the Wigner quasiprobability of a rotated density operator 
about an angle $-\varphi$ that means to $R(-\varphi)\varrho \big(R(-\varphi)
\big)^\dagger$ and according to Eqs.(3.10), (3.12), (3.14) or the definition
of the Radon transform in Eq.(2.2) one obtains
\begin{equation}   
\langle q;\varphi|\varrho|q;\varphi\rangle = \int_{-\infty}^{+\infty}dp\,
W(q\cos\varphi-p\sin\varphi,q\sin\varphi+p\cos\varphi) = \breve{W}
(\cos\varphi,\sin\varphi;q).
\end{equation}
Thus the relation (4.1) is proved. From the well-known position 
representation $\langle q|n\rangle$ of the Fock states $|n\rangle$ it follows
\begin{equation}
\langle q;\varphi|n\rangle = \langle q|R(-\varphi)|n \rangle = 
\frac{\exp(-in\varphi)}{(\hbar \pi)^{\frac{1}{4}}}\exp\left(-\frac{q^2}
{2\hbar}\right) \frac{1}{\sqrt{2^n n!}}H_{n}\left(\frac{q}{\sqrt{\hbar}}
\right).
\end{equation}
With the summation formula of Mehler \cite{bat} 
\begin{equation}
\sum_{n=0}^{\infty}\frac{z^n}{2^n n!}H_n(x)H_n(y) =\frac{1}{\sqrt{1-z^2}}
\exp\left\{\frac{2xyz-\left(x^2+y^2\right)z^2}{1-z^2}\right\},
\end{equation}
which is not difficult to prove \cite{w2}
one calculates by Fock-state expansion the more general scalar product 
( Green's function to the wave equation for the harmonic oscillator )
\begin{equation}
\langle q;\varphi|q';\varphi'\rangle = \frac{1}{\sqrt{\hbar\pi\left(
1-\exp\{-i2(\varphi-\varphi')\}\right)}}\exp\left\{i\frac{\left(q^2+q'^{\,2}
\right)\cos(\varphi-\varphi') -2qq'}{2\hbar\sin(\varphi-\varphi')}\right\}.
\end{equation}
In the limiting case $\varphi'\to \varphi$ one obtains the delta function 
$\delta(q-q')$ and in case of $\varphi'=\varphi+\pi/2$ the scalar product 
$\langle q|p\rangle$ if one substitutes $q'\equiv p$.

   The Wigner quasiprobability contains the complete information about
the state described by the density operator. The reconstruction of the 
density operator from the Wigner quasiprobability can be made by 
transition operators. As was already mentioned the Wigner quasiprobability 
has the unique property among all quasiprobabilities that the transition 
operators for the transition from the density operator to the 
quasiprobability by forming the trace and from the quasiprobability back to 
the density operator by a phase-space integral are the same. The 
reconstruction formula of the density operator from the Wigner 
quasiprobability by means of these transition operators $T_0(q,p)$ in real 
or $T_0(\alpha,\alpha^*)$ in complex representation has the following form
\begin{eqnarray}
&&\varrho=2\hbar\pi \int dq\wedge dp \,W(q,p) T_0(q,p) = \pi \int \frac{i}{2}
d\alpha \wedge d\alpha^* \,W(\alpha,\alpha^*) T_0(\alpha,\alpha^*),
\nonumber\\&& W(q,p)=\langle \varrho T_{0}(q,p)\rangle,\qquad W(\alpha,
\alpha^*)=\langle \varrho T_{0}(\alpha,\alpha^*)\rangle.
\end{eqnarray}
The transition operator for the Wigner quasiprobability is essentially  
the displaced parity operator [64--66,38].
There exist many 
representations for this transition operator. We use for our purpose the 
representation explicitly contained in Eq.(3.1). The reconstruction 
formula (4.8) can be written then in real representation in the following
form  
\begin{eqnarray} 
\varrho\!\! &=&\!\! 2\hbar\pi \int dq \wedge dp \,W(q,p) \exp\left(
-Q\frac{\partial}{\partial q}-P\frac{\partial}{\partial p}\right) 
\,\delta(q)\delta(p) \nonumber\\  &=&\!\! 2\hbar\pi \int dq \wedge dp \,
\delta(q)\delta(p) \exp\left( Q\frac{\partial}{\partial q}+P\frac{\partial}
{\partial p}\right) W(q,p) \nonumber\\ &=&\!\! 2\hbar \pi 
\sum_{k=0}^{\infty} \sum_{l=0}^{\infty} \frac{{\cal{S}}\{Q^k P^l\}}{k!l!}
\left\{\frac{\partial^{k+l}}{\partial q^k \partial p^l}
W(q,p)\right\}_{(q=0,p=0)}.
\end{eqnarray}
where $\cal{S}\{\ldots\}$ means symmetrical ordering of the content in braces.
This formula shows first of all that the Wigner quasiprobability as all other 
quasiprobabilities too contains a great redundancy and that for the 
reconstruction of the density operator it is already sufficient to know
the Wigner quasiprobability $W(q,p)$ in an arbitrarily small neighbourhood
of $(q=0,p=0)$. One could introduce in Eq.(4.9) the representation of the
Wigner quasiprobability $W(q,p)$ by its Radon transform $\breve{W}(u,v;q)$
explicitly given in Eq.(2.13) and has a representation of the density 
operator by the rotated quadrature components. Such a form, however, is 
not very convenient, for example, for the calculation of the matrix 
elements of the density operator in the Fock-state representation. 
It is better for some purposes to use the Fourier transform
$\tilde{W}(u,v)$ of the Wigner quasiprobability as an intermediate step.
 
   The introduction of the Fourier transform of the Wigner quasiprobability 
according to Eq.(2.7) into Eq.(4.9) yields
\begin{eqnarray}
\varrho\!\!&=&\!\! 2\hbar \pi \int dq\wedge dp \,\delta(q)\delta(p)
\exp\left(Q\frac{\partial}{\partial q}+ P\frac{\partial}{\partial p}\right)
\frac{1}{(2\pi)^2} \int du\wedge dv \tilde{W}(u,v)\exp\left\{i(uq+vp)\right\} 
\nonumber\\&=&\!\!\frac{\hbar}{2\pi}\int du \wedge dv \exp\left\{i(Qu+Pv)
\right\} \tilde{W}(u,v).
\end{eqnarray}
Now, by expressing the Fourier transform of the Wigner quasiprobability  
by the Radon transform of the Wigner quasiprobability one finds ( $I$ is
the identity operator )
\begin{equation}
\varrho = \frac{\hbar}{2\pi} \int_{-\infty}^{+\infty} dc \int du \wedge dv
\exp\{-i(Ic-Qu-Pv)\}\breve{W}(u,v;c).
\end{equation}
After the substitutions
\begin{equation}
u\equiv r \cos\varphi\quad v\equiv r\sin\varphi,\quad c\equiv rq,
\end{equation}
one arrives at ( note the general identity $ \int_{0}^{+\infty}dr \,r
\int_{0}^{2\pi}f(r,\varphi)=\int_{-\infty}^{+\infty} dr |r| 
\int_{0}^{\pi} f(r,\varphi)$ for unique functions of $r$ and $\varphi$ 
over the plane )
\begin{eqnarray}
\varrho \!\!&=&\!\! \frac{\hbar}{2\pi} \int_{0}^{\pi} d\varphi 
\int_{-\infty}^{+\infty} dq \int_{-\infty}^{+\infty} dr|r| 
\exp\{-ir(Iq-Q\cos\varphi -P \sin\varphi)\}
\breve{W}(\cos\varphi,\sin\varphi;q) \nonumber\\ &=&\!\! \frac{1}{\pi}
\int_{0}^{\pi} d\varphi \int_{-\infty}^{+\infty} dq \,\breve{W}(\cos\varphi,
\sin\varphi;q) \,\hbar\, \frac{\partial}{\partial q}\left( {\cal{P}}\frac{1}
{Iq-Q(\varphi)} \right) \nonumber\\ &=&\!\! 
- \frac{1}{\pi}\int_{0}^{\pi} d\varphi \int_{-\infty}^{+\infty} dq \,
\left({\cal{P}}\frac{1}{Iq-Q(\varphi)} \right) \,\hbar\,
\frac{\partial}{\partial q} \breve{W}(\cos\varphi,\sin\varphi;q),
\end{eqnarray}
where $Q(\varphi)$ is the rotated operator $Q$ according to Eq.(3.10). 
Recall that ${\cal{P}}(1/x)\equiv{\cal{R}}(1/x)$ means the principal value of 
$1/x$ and its first derivative $-{\cal{R}}(1/x^2)$ the canonical 
regularization of $-1/x^2$ as explained in section 2, Eq.(2.15). However,
the operator character of the singularities in this formula makes it more
difficult to interpret them. A possible way to do this is to make a Taylor
series expansion in powers of $Q(\varphi)$ in the following way
\begin{eqnarray}
\varrho \!\!&=&\!\! 
-\frac{1}{\pi}\int_{0}^{\pi} d\varphi 
\int_{-\infty}^{+\infty}dq \,\hbar 
\breve{W}(\cos\varphi,\sin\varphi;q) 
{\cal R} \frac{1}{\big(Iq-Q(\varphi)\big)^2} \nonumber\\ &=&\!\!
-\sum_{k=0}^{\infty}(k+1) \frac{1}{\pi}\int_{0}^{\pi} d\varphi \,
\Big(Q(\varphi)\Big)^k \int_{-\infty}^{+\infty} dq \,\hbar 
\breve{W}(\cos\varphi,\sin\varphi;q) {\cal R}\frac{1}{q^{2+k}}
\nonumber\\&=& \!\!
-\sum_{k=0}^{\infty}(k+1) \frac{1}{\pi}\int_{0}^{\pi} d\varphi \,
\Big(Q(\varphi)\Big)^k \nonumber\\ &&\!\!
\times \int_{0}^{+\infty} dq \,\frac{\hbar}{q^{2+k}}
\Bigg\{\breve{W}(\cos\varphi,\sin\varphi;q) 
+(-1)^k \breve{W}(\cos\varphi,\sin\varphi;-q) \nonumber\\ && \!\!
-\sum_{l=0}^k \left(1+(-1)^{k+l}\right)\,\frac{q^l}{l!}\,
\frac{\partial^{\,l} \breve{W}}{\partial q^{l}}(\cos\varphi,\sin\varphi;0) 
\Bigg\}.
\end{eqnarray}
In the last part of this representation we have explicitly written down the 
meaning of the canonical regularization of powers $1/q^{2+k}$ ( see also 
\cite{shilov} ). The integration over $q$ goes here only from zero to 
plus infinity. We wrote this equation in detail because the content of 
formulae such as (4.13) was not represented with clarity in the literature 
about quantum tomography. However, it seems that Eqs.(4.14) does 
not provide a very convenient approach and mostly it is better to make the
integration with respect to $r$ in Eq.(4.13) not before calculating 
the matrix elements of the density operator in Fock-state representation or
in other representation. Nevertheless, it cannot be excluded that sometimes 
some initial terms of the Taylor series expansion in (4.14) could give 
already a good approximation that must be investigated but the same 
approximations can be obviously obtained by Taylor series expansion of the 
exponentials in (4.13) where the integration over $r$ is accomplished in a 
later stage of the calculations.

A more convenient form of the reconstruction formula of the density 
operator from its Radon transform for many purposes can be obtained when
the integration with respect to $r$ is not accomplished before making the
transition to normal ordering. Our method is similar to the method used in
\cite{paul3,ariano}. In such a way by transition to normal ordering one 
obtains 
\begin{eqnarray}
\varrho \!\!&=&\!\! \frac{1}{\pi} \int_{0}^{\pi} d\varphi 
\int_{-\infty}^{+\infty} dq \,\breve{W}(\cos\varphi,\sin\varphi;q)
\frac{\partial}{\partial q}\left(\,i\,\frac{\hbar}{2}\,\int_{-\infty}^{+
\infty}dr\;\frac{r}{|r|}\exp\left(-ir(Iq- Q(\varphi) \right) \right) 
\nonumber\\&=&\!\! \frac{1}{\pi} \int_{0}^{\pi} d\varphi 
\int_{-\infty}^{+\infty} dq \,\breve{W}(\cos\varphi,\sin\varphi;q)
\nonumber\\ &&\!\! \times \frac{\partial}{\partial q}  
\left\{ R(\varphi)\,
i\frac{\hbar}{2}\int_{-\infty}^{+\infty} dr\,\frac{r}{|r|}\exp\left(-irq-
r^2\frac{\hbar}{4}\right) \exp\left(ir\sqrt{\frac{\hbar}{2}}a^\dagger\right)
\exp\left(ir\sqrt{\frac{\hbar}{2}}a\right) 
\left(R(\varphi)\right)^\dagger \right\},\nonumber\\
\end{eqnarray}
and by the substitution of $r\rightarrow i\partial/\partial q$ in 
parts of the integral over $r$ 
\begin{eqnarray}
\varrho\!\! &=&\!\!
\frac{1}{\pi} \int_{0}^{\pi} d\varphi 
\int_{-\infty}^{+\infty} dq \,\breve{W}(\cos\varphi,\sin\varphi;q)
\nonumber\\ &&\!\! \times \sqrt{\hbar}\frac{\partial}{\partial q}  
\Bigg\{ R(\varphi)
\exp\left(-a^\dagger\sqrt{\frac{\hbar}{2}}\frac{\partial}{\partial q}
\right)\exp\left(-a\sqrt{\frac{\hbar}{2}}\frac{\partial}{\partial q}\right)
\left(R(\varphi)\right)^\dagger  
\nonumber\\ &&\!\!\times 
\:i\,\frac{\sqrt{\hbar}}{2}\int_{-\infty}^{+\infty} dr\,\frac{r}{|r|}
\exp\left(-iqr-\frac{\hbar}{4}r^2\right) \Bigg\}.
\end{eqnarray}
We obtained a representation where a normally ordered operator part is 
separated from an integral onto which it acts. Now, we make the variable
substitution 
\begin{equation}
x \equiv \frac{q}{\sqrt{\hbar}},\quad \frac{\partial}{\partial x} = 
\sqrt{\hbar}\frac{\partial}{\partial q},\quad dx = \frac{dq}{\sqrt{\hbar}}.
\end{equation}
The integral over $r$ in Eq.(4.16) can be transformed and solved in the 
following way
\begin{eqnarray}
&&\!\!i\frac{\sqrt{\hbar}}{2}\int_{-\infty}^{+\infty} dr\,\frac{r}{|r|}
\exp\left(-i\sqrt{\hbar}xr-\frac{\hbar}{4}r^2\right) = 
2\exp\left(-x^2\right) \int_0^x dt\,\exp\left(t^2 \right) \equiv 2 F(x)
\nonumber\\ &=&\!\! \sqrt{2} D_0\left(\sqrt{2}x\right) \frac{i}{2}
\left\{ D_{-1}\left(i\sqrt{2}x\right)-D_{-1}\left(-i\sqrt{2}x\right)\right\}
\equiv h_0(x) g_0(x),
\end{eqnarray}
with the abbreviations
\begin{eqnarray}
&& h_{0}(x)\equiv\frac{1}{\pi^{\frac{1}{4}}}\exp\left(-\frac{x^2}{2}\right)
=\frac{1}{\pi^{\frac{1}{4}}} D_0\left(\sqrt{2}x\right),\quad 
\int_{-\infty}^{+\infty}dx\,\left(h_0(x)\right)^2=1, \nonumber\\ &&
g_{0}(x)\equiv 2\pi^{\frac{1}{4}} \exp\left(-\frac{x^2}{2}\right) \int_0^x dt
\,\exp\left(t^2\right)= \sqrt{2}\pi^{\frac{1}{4}} \frac{i}{2}
\left\{ D_{-1}\left(i\sqrt{2}x\right)-D_{-1}\left(-i\sqrt{2}x\right)\right\},
\nonumber\\ && W\big(h_0(x),g_0(x)\big) = 2,\qquad W(f(x),g(x))\equiv 
f(x)g^{(1)}(x)-f^{(1)}(x)g(x),
\end{eqnarray}
where $W\left(f(x),g(x)\right)$ denotes the Wronskian of two functions 
$f(x)$ and $g(x)$ and $D_{\nu}(z)$ the functions of the parabolic cylinder, 
in particular
\begin{equation}
D_{-1}\left(\pm i\sqrt{2}x\right)= \sqrt{\frac{\pi}{2}}
\exp\left(-\frac{x^2}{2}\right)\Big( 1 \mp {\mit \Phi}(ix) \Big), \quad 
{\mit \Phi}(\mu x)\equiv \frac{2\mu}{\sqrt{\pi}}
\int_0^{x}ds\,\exp\left(-\mu^2 s^2\right),
\end{equation}
with ${\mit \Phi}(z)$ as the error function. The error function of imaginary
argument has a close relation to the Dawson integral denoted by $F(z)$ in 
(4.18).
The function $h_{0}\left(q/\sqrt{\hbar}\right)/\hbar^{1/4}$ is the normalized 
solution of the wave equation for the harmonic oscillator in the ground
state that means the eigenstate of the number operator to the eigenvalue
$n=0$ in "position" representation. It possesses even parity. 
The function $ g_{0}\left(q/\sqrt{\hbar}\right)$ is under all nonnormalizable 
eigenfunctions of the number operator to the eigenvalue $n=0$ the 
eigenfunction with odd parity.

The normally ordered operator part in Eq.(4.16) is convenient for the 
calculation of matrix elements in the Fock-state representation and in the
coherent-state representation. We begin with the simpler last case.
Due to the relation
\begin{equation}
\langle \alpha|R(\varphi) \exp\left(- \frac{a^\dagger}{\sqrt{2}}
\frac{\partial}{\partial x}\right)\exp\left(-\frac{a}{\sqrt{2}} 
\frac{\partial}{\partial x}\right)\!\left(R(\varphi)\right)^\dagger 
|\alpha \rangle \nonumber\\ = 
\exp\left(-\frac{\alpha {\rm e}^{-i\varphi}+
\alpha^* {\rm e}^{i\varphi}}{\sqrt{2}}\frac{\partial}{\partial x} \right),
\end{equation}
it acts onto the integral in (4.16) as a displacement operator of the 
argument and one obtains from (4.16) immediately the following reconstruction
of the coherent-state quasiprobability from the Radon transform of the Wigner
quasiprobability
\begin{eqnarray}
Q(\alpha,\alpha^*)\!\!&\equiv&\!\!
\frac{\langle \alpha|\varrho|\alpha \rangle}{\pi}\nonumber\\ \!\!&=&\!\!
\frac{1}{\pi} \int_{0}^{\pi} d\varphi 
\int_{-\infty}^{+\infty} dq \,\breve{W}(\cos\varphi,\sin\varphi;q)\,
\frac{2\sqrt{\hbar}}{\pi} \frac{\partial}{\partial q} F\left(\frac{q}
{\sqrt{\hbar}}-\frac{\alpha {\rm e}^{-i\varphi}+\alpha^* {\rm e}^{i\varphi}}
{\sqrt{2}}\right).\nonumber\\
\end{eqnarray}
Herein, $F(z)$ denotes again the Dawson integral which definition can be 
taken from Eq.(4.18).

We now consider the reconstruction of the Fock-state matrix elements. First
of all, we find for the Fock-state matrix elements of the operator part in 
(4.16)
\begin{eqnarray}
S_{m,n}\left(\frac{\partial}{\partial x}\right)\!\!&\equiv&\!\!
\langle m|R(\varphi) \exp\left(-\frac{a^\dagger}{\sqrt{2}}
\frac{\partial}{\partial x}\right)\exp\left(-\frac{a}{\sqrt{2}} 
\frac{\partial}{\partial x}\right)\left(R(\varphi)\right)^\dagger 
|n \rangle 
\nonumber\\ &=&\!\! 
\frac{\exp\big\{i(m-n)\varphi\big\}}{\sqrt{m!n!}} \sum_{j=0}^{\{m,n\}} 
\frac{m!n!}{j!(m-j)!(n-j)!}\left(-\frac{1}{\sqrt{2}}\frac{\partial}
{\partial x}\right)^{m+n-2j} \nonumber\\ &=&\!\!
\exp\big\{i(m-n)\varphi\big\} \sqrt{\frac{n!}{m!}}\left(-\frac{1}{\sqrt{2}}
\frac{\partial}{\partial x}\right)^{m-n}L_{n}^{m-n}\left(-\frac{1}{2}
\frac{\partial^2}{\partial x^2}\right),
\end{eqnarray}
where $L_{n}^{m-n}(z)$ denote the associated Laguerre polynomials.
By forming the Fock-state matrix elements in (4.16) and by inserting then 
(4.23) one obtains the following structure of the reconstruction formula 
\begin{equation}
\langle m|\varrho|n\rangle =
\frac{1}{\pi} \int_{0}^{\pi} d\varphi \int_{-\infty}^{+\infty} dq \,
\breve{W}(\cos\varphi,\sin\varphi;q)\,\exp\big\{i(m-n)\varphi\big\} 
F_{m,n}\left(\frac{q}{\sqrt{\hbar}}\right),
\end{equation}
with functions $F_{m,n}(x)$ defined in the following way
\begin{eqnarray}
F_{m,n}(x)\!\!&\equiv&\!\! \frac{\partial}{\partial x} \left\{ \frac{1}
{\sqrt{m!n!}} \sum_{j=0}^{\{m,n\}} \frac{m!n!}{j!(m-j)!(n-j)!}
\left(-\frac{1}{\sqrt{2}}\frac{\partial}{\partial x}\right)^{m+n-2j} 
h_0(x)g_0(x) \right\}\nonumber\\ \!\!&=&\!\!
\frac{1}{\sqrt{m!n!}} \sum_{j=0}^{\{m,n\}} \frac{m!n!}{j!(m-j)!(n-j)!}
\left(-\frac{1}{\sqrt{2}}\frac{\partial}{\partial x}\right)^{m+n-2j} 
F_{0,0}(x)=F_{n,m}(x).
\end{eqnarray}
The functions $\exp\big\{i(m-n)\varphi\big\}F_{m,n}(x)$ are called  
"pattern functions" for the reconstruction of the Fock-state matrix elements 
[12--14]. 
However, contrary to the cited papers we have 
defined them with a factor $\pi$ larger because then the integral over the 
angle $\varphi$ in (4.24) with $1/\pi$ in front can be considered as an 
avering over the angle. The angle-dependent part is explicitly splitted 
in (4.24) from the more complicated position-dependent part. The pattern 
functions $F_{m,n}(x)$ are symmetrical in the indices by definition (4.25).

   Let us consider the explicit representation of the pattern functions by 
functions of the parabolic cylinder $D_{\nu}(z)$. Due to the relations
\begin{equation}
\left(\pm\frac{i}{\sqrt{2}}\left(x-\frac{\partial}{\partial x}\right)
\right)^n D_{\nu}\Big(\pm i\sqrt{2}x\Big)= \frac{{\mit \Gamma}(\nu+1)}
{{\mit \Gamma}(\nu-n+1)}D_{\nu-n}\Big(\pm i\sqrt{2}x\Big),
\end{equation}
following from basic definitions of these functions one obtains by 
accomplishing the differentiations in (4.25) and taking into account (4.18) 
and (4.19)
\begin{eqnarray}
F_{m,n}(x)\!\!&=&\!\! \exp\left(-\frac{x^2}{2}\right) 2\sqrt{m!n!}
\sum_{j=0}^{\{m,n\}}\frac{(-1)^j (m+n+1-2j)!}{j!(m-j)!(n-j)!}\nonumber\\
&&\!\! \times \frac{i^{m+n}}{2}\left\{D_{-(m+n+2-2j)}\Big(i\sqrt{2}x\Big)
+(-1)^{m+n} D_{-(m+n+2-2j)}\Big(-i\sqrt{2}x\Big) \right\}.
\end{eqnarray}
This is near to the representation derived in \cite{paul3} but the 
calculations are made there more generally for the reconstruction of the 
matrix elements in the Fock-state basis from the Radon transform of the
$s$-ordered quasiprobabilities. In section 6 we will obtain an alternative  
representation of the pattern functions in form of a series over even or
odd Hermite polynomials.

   We now derive a specific structure of the pattern functions as derivatives
of the products of a normalizable with a nonnormalizable wave function of the
harmonic oscillator 
[18-21].
The operator in (4.23) acts onto the product $h_0(x)g_0(x)$ as an excitation
operator. The creation operator $a^\dagger$ can be substituted in application 
to wave functions $\langle q|\psi \rangle \equiv \psi(q)$ by
\begin{equation}
a^\dagger =\frac{Q-iP}{\sqrt{2\hbar}} \longrightarrow \frac{1}{\sqrt{2}}
\left(\frac{q}{\sqrt{\hbar}}-\sqrt{\hbar}\frac{\partial}{\partial q}\right) = 
\frac{1}{\sqrt{2}}\left(x-\frac{\partial}{\partial x}\right),\quad x\equiv 
\frac{q}{\sqrt{\hbar}}.
\end{equation}
Therefore, the higher excitations of the eigenfunctions of the number 
operator can be defined by
\begin{eqnarray}
&&\left(|n\rangle, |n\rangle'\right) \equiv \frac{a^{\dagger\,n}}{\sqrt{n!}}
\left(|0\rangle, |0\rangle'\right) \longrightarrow \left(h_n(x),g_n(x)\right)
\equiv \frac{1}{\sqrt{2^n n!}}\left(x-\frac{\partial}{\partial x}\right)^n 
\left(h_0(x),g_0(x)\right), \nonumber\\ && \int_{-\infty}^{+\infty}dx\,h_m(x)
h_n(x)=\delta_{m,n},\quad W(h_{n}(x),g_{n}(x))=2,
\end{eqnarray}
where $|n\rangle'$ is the abstract notation for the nonnormalizable 
eigenstates of the number operator with parity $(-1)^{n+1}$.
The functions $h_n(x)$ are the orthonormalized Hermite functions or with the
substitution $x=q/\sqrt{\hbar}$ and the absent normalization factor 
$1/\hbar^{(1/4)}$ the wave functions $\langle q|n\rangle$ given in (4.5).
In the same way, the annihilation operator $a$ in "position" representation
can be represented according to
\begin{equation}
a=\frac{Q+iP}{\sqrt{2\hbar}} \longrightarrow 
\frac{1}{\sqrt{2}}\left(
\frac{q}{\sqrt{\hbar}}+\sqrt{\hbar}\frac{\partial}{\partial q}\right) = 
\frac{1}{\sqrt{2}}\left(x+\frac{\partial}{\partial x}\right).
\end{equation}
However, there is a very important difference concerning the action of the
operator $a$ to the vacuum state $|0\rangle$ and to the state $|0\rangle'$,
in "position" representation
\begin{eqnarray}
&&\frac{1}{\sqrt{2}}\left(x+\frac{\partial}{\partial x}\right)h_{0}(x)=0, 
\quad \frac{1}{\sqrt{2}}\left(x+\frac{\partial}{\partial x}\right)g_{0}(x) 
=\pi^{\frac{1}{4}}\sqrt{2}\exp\left(\frac{x^2}{2}\right)\equiv g_{-1}(x),
\nonumber\\&& 
\frac{1}{\sqrt{2}}\left(x-\frac{\partial}{\partial x}\right)g_{-1}(x)=0,\quad 
\frac{1}{\sqrt{2}}\left(x-\frac{\partial}{\partial x}\right)
\frac{1}{\sqrt{2}}\left(x+\frac{\partial}{\partial x}\right)g_{-1}(x)=
-g_{-1}(x).\nonumber\\
\end{eqnarray}
Obviously, by repeated action of the annihilation operator in position 
representation onto the state $g_{-1}(x)$ one can obtain nonnormalizable
eigenstates $g_{-n}(x)$ of the number operator to the eigenvalues $-n$
whereas for the normalizable eigenstates the corresponding series is 
truncated below the vacuum state $h_{0}(x)$. We do not investigate this 
here more in detail but show that the asymmetry in the action of the 
annihilation operator onto $h_{0}(x)$ and $g_{0}(x)$ affects an assymmetry
in the representation of the pattern functions by the series $h_{m}(x)$ and
$g_{n}(x)$.

We now prove that 
\begin{equation}
F_{m,n}(x)=\frac{\partial}{\partial x}\left\{h_{m}(x)g_{n}(x)\right\}, 
\quad ( m \le n+1 ),\qquad (\;F_{m,n}(x)=F_{n,m}(x)\;). 
\end{equation}
The proof of this formula can be made by complete induction using the 
recursion relations
\begin{eqnarray}
&&S_{m,n+1}(u)=\frac{1}{\sqrt{n+1}}\left\{
-{\rm e}^{-i\varphi} \frac{u}{\sqrt{2}}S_{m,n}(u) 
+\sqrt{n}S_{m-1,n}(u)\right\},\nonumber\\&&
S_{m+1,n}(u)=\frac{1}{\sqrt{m+1}}\left\{
-{\rm e}^{i\varphi} \frac{u}{\sqrt{2}}S_{m,n}(u)
+\sqrt{n}S_{m,n-1}(u)\right\},
\end{eqnarray}
following from (4.23). 
The relation (4.32) is obviously true for $m=0$ and arbitrary $n$ because of
\begin{equation}
\frac{\partial}{\partial x}
\frac{1}{\sqrt{2^n n!}}\left(-\frac{\partial}{\partial x}\right)^n 
\left\{h_0(x)g_0(x)\right\} = \frac{\partial}{\partial x}
h_0(x)\frac{1}{\sqrt{2^n n!}}
\left(x-\frac{\partial}{\partial x}\right)^n g_0(x)  = 
\frac{\partial}{\partial x}\{h_0(x)g_n(x)\}.
\end{equation}
On the other side, however, one finds
\begin{eqnarray}
&&\frac{\partial}{\partial x}
\frac{1}{\sqrt{2^11!}}\left(-\frac{\partial}{\partial x}\right) 
\left\{h_0(x)g_0(x)\right\} = \frac{\partial}{\partial x}
\left\{h_{1}(x)g_{0}(x)-\sqrt{2}\right\},\nonumber\\ &&
\frac{\partial}{\partial x}
\frac{1}{\sqrt{2^2 2!}}\left(-\frac{\partial}{\partial x}\right)^2 
\left\{h_0(x)g_0(x)\right\} = \frac{\partial}{\partial x}
\left\{h_{2}(x)g_{0}(x)-\sqrt{2}x\right\}.
\end{eqnarray}
This means that (4.32) is not generally true for all $(m,n=0)$ without 
restriction. Let us give the proof of (4.32) by complete induction. 
Suppose that it is true for a certain $(m,n)$. Then by using the recursion
relation (4.33) one can prove without difficulties that it is also true
for $(m,n+1)$. We do not explicitly write this down because we will write
down the induction from $(m,n)$ to $(m+1,n)$ which is very similar but 
shows in addition a specific difficulty. By using (4.33) one finds by 
induction from $(m,n)$ to $(m+1,n)$
\begin{eqnarray}
\!\!\!\!\!&&\!\! \frac{\partial}{\partial x}
S_{m+1,n}\left(\frac{\partial}{\partial x}\right)
\left\{h_0(x)g_0(x)\right\} \nonumber\\ \!\!\!\!&=&\!\! 
\frac{\partial}{\partial x}
\frac{{\rm e}^{i(m+1-n)\varphi}}{\sqrt{m+1}}\Bigg\{
g_n(x)\frac{1}{\sqrt{2}}\left(x-\frac{\partial}{\partial x}\right)h_m(x)
-h_m(x) \Bigg(\frac{1}{\sqrt{2}}\left(x+\frac{\partial}{\partial x}\right)
g_n(x) -\sqrt{n} g_{n-1}(x)\Bigg) \Bigg\}
\nonumber\\[2mm] \!\!\!\!&=&\!\! {\rm e}^{i(m+1-n)\varphi}
\frac{\partial}{\partial x}
\left\{h_{m+1}(x)g_n(x)-\delta_{n,0}\,h_m(x)
g_{-1}(x) \right\},
\end{eqnarray}
where the relations (4.29) and (4.30) are applied. 
The sum term proportional to $\delta_{n,0}$ in brackets which have its
origin in the relations (4.31) prevents the full proof of (4.32) for all 
$(m+1,n)$. However, if one raises at once $n$ to $n+1$ that means if one 
makes the complete induction from $(m,n)$ to $(m+1,n+1)$ then the proof makes no 
difficulties because of $\delta_{n+1,0}=0$ for all nonnegative $n$. This 
means that the first relation in Eq.(4.32) is proved with the restriction 
$m\le n+1$ because it is right for $(m=0,n)$ and $(m=1,n)$. 
If we do not consider the restriction $m\le n+1$ then one obtains 
new pattern functions by the first of the relations in Eq.(4.32) which do not 
coincide with $F_{m,n}(x)$ defined by (4.25) or (4.27) for $m>n+1$ but can 
be used by full rights as equivalent pattern functions. We investigate this 
nonuniqueness of the pattern functions for the reconstruction of the 
Fock-state matrix elements furthermore in sections 6 and 7. The pattern 
functions given by (4.25) or (4.27) possess the advantage that they vanish 
in infinity.

\setcounter{chapter}{5}
\setcounter{equation}{0}
\chapter*{5. Reconstruction of the normally ordered moments from the Radon
transform of the Wigner quasiprobability} 
 
   The normally ordered moments $\langle a^{\dagger\,k}a^l \varrho\rangle$
can be directly reconstructed from the Radon transform of the Wigner 
quasiprobability without the intermediate calculation of the matrix elements 
in the Fock-state representation as found by Richter \cite{ri2}. A very 
simple derivation of this reconstruction is given in \cite{w7} where is also 
shown that the integration over the angle in the Radon transform made in 
\cite{ri2} is not necessary and can be substituted by certain summations over 
discrete angles that leads to more basic formulas.

The starting point for the reconstruction of the normally ordered moments
is the following representation of the density operator by the normally 
ordered moments derived in [63--65]
\begin{equation}
\varrho = \sum_{k=0}^{\infty} \sum_{l=0}^{\infty} a_{k,l}
\langle a^{\dagger\,k}a^l \varrho \rangle , \quad a_{k,l}\equiv 
\sum_{j=0}^{\{k,l\}}\frac{(-1)^j|l-j\rangle \langle k-j|}
{j!\sqrt{(k-j)!(l-j)!}}.
\end{equation}
It can be easily obtained from the normally ordered expansion of the 
basic Fock-state operators $|m\rangle\langle n|$ in powers of the boson 
operators \cite{lou}. For the Radon transform of the Wigner quasiprobability
in the form $\breve{W}(\cos\varphi,\sin\varphi;q)$ one finds from (5.1)
\begin{equation}
\breve{W}(\cos\varphi,\sin\varphi;q)=
\langle q;\varphi|\varrho|q;\varphi \rangle =
\sum_{k=0}^{\infty}\sum_{l=0}^{\infty}\langle q;\varphi|a_{k,l}|q;\varphi 
\rangle \langle  a^{\dagger \,k}a^l \varrho \rangle,
\end{equation}
and with the explicit form of the position representation of the number states
by Hermite functions 
\begin{eqnarray}
\langle q;\varphi|a_{k,l}|q;\varphi \rangle \!\!\!&=& \!\!
\frac{1}{\sqrt{\pi\hbar}}
\exp\left(\!-\frac{q^2}{\hbar}\right)\frac{e^{i(k-l)\varphi}}
{\sqrt{2^{k+l}}k!l!}\sum_{j=0}^{\{k,l\}}\frac{(-2)^j k!l!}{j!(k-j)!(l-j)!}
H_{k-j}\left(\frac{q}{\sqrt{\hbar}}\right) H_{l-j}\left(\frac{q}{\sqrt{\hbar}}
\right) \nonumber\\ &=&\!\!\!
\frac{1}{\sqrt{\pi\hbar}}
\exp\left(\!-\frac{q^2}{\hbar}\right)\frac{e^{i(k-l)\varphi}}
{\sqrt{2^{k+l}}k!l!} H_{k+l}\left(\frac{q}{\sqrt{\hbar}}\right).
\end{eqnarray}
Here we used a known identity for finite sums over products of Hermite 
polynomials ( see Eq.10.13. (36) in \cite{bat} ) which can be proved by 
complete induction. By multiplication of Eq.(5.3) by 
$H_{n}\left(q/\sqrt{\hbar}\right)$ and by using the well-known completeness 
relations for the Hermite functions one finds
\begin{eqnarray}
&&\!\!\!\frac{1}{\sqrt{2^n}}\int_{-\infty}^{+\infty}dq \,
\breve{W}(\cos\varphi,\sin\varphi;q)H_{n}\left(\frac{q}{\sqrt{\hbar}}\right)=
\sum_{k=0}^n \frac{n!}{k!(n-k)!}e^{i(2k-n)\varphi}\langle a^{\dagger\,k}
a^{n-k} \varrho\rangle \nonumber\\ &=&\!\!\!\Big\langle {\cal{N}}
\Big\{\Big(a e^{-i\varphi}+a^\dagger e^{i\varphi}\Big)^n \Big\}\varrho 
\Big\rangle = \left(\sqrt{\frac{2}{\hbar}}\:\right)^{\!n} \big\langle 
{\cal{N}} \big\{\big(Q(\varphi)\big)^n\big\}\varrho \big\rangle.
\end{eqnarray}
The symbol ${\cal N}\{\ldots\}$ denotes normal ordering of the content in
braces. The system of equations (5.4) for the normally ordered moments 
$ \langle a^{\dagger\,k}a^{n-k}\varrho \rangle $ can be solved by using the 
properties of the solutions of the circle division problem. The solutions $z$
for the division of the unit circle in $n+1$ equal sectors 
( harmonic division )
\begin{equation}
0=z^{n+1}-1=(z-1)\sum_{m=0}^{n}z^m,\quad z=\exp\left(is\frac{2\pi}{n+1}\right),\quad 
s=0,1,\ldots,n ,
\end{equation}
possess the property 
\begin{equation}
\sum_{m=0}^{n}z^m= \sum_{m=0}^{n}\exp\left(ims\frac{2\pi}{n+1}\right)=
(n+1)\,\delta_{s,0}.
\end{equation}
By using these orthogonality relations of the solutions of the circle 
division problem one finds from Eq.(5.4)
\begin{eqnarray}
\langle  a^{\dagger \,k}a^l \varrho \rangle \!\!&=&\!\!\frac{k!l!}
{(k+l+1)!} \sum_{m=0}^{k+l}\exp\left\{-i(k-l)\left(\varphi_{0}+\frac{m\pi}
{k+l+1}\right)\right\}\frac{1}{\sqrt{2^{k+l}}} \nonumber\\ &&\!\!\times
\int_{-\infty}^{+\infty}dq\,\breve{W}\left\{\cos\left(\varphi_{0}+\frac{m\pi}
{k+l+1}\right),\sin\left(\varphi_{0}+\frac{m\pi}{k+l+1}\right);q\right\}
H_{k+l}\left(\frac{q}{\sqrt{\hbar}}\right).\nonumber\\
\end{eqnarray}
Herein, $\varphi_{0}$ denotes an arbitrary initial angle. This formula is the
only place we know where our notations of the Radon transform seem to be a
little bulky. By integration over the arbitrary initial angle $\varphi_{0}$
one obtains \cite{ri2,w7}
\begin{equation}
\langle  a^{\dagger \,k}a^l \varrho \rangle = \frac{k!l!}{(k+l)!}        
\frac{1}{\pi}\int_{0}^{\pi} \! d\varphi\,\exp\left\{-i(k-l)
\varphi\right\} \frac{1}{\sqrt{2^{k+l}}}\int_{-\infty}^{+\infty} \! dq\,
\breve{W}\left(\cos\varphi,\sin\varphi;q\right) H_{k+l}\left(\frac{q}
{\sqrt{\hbar}}\right).
\end{equation}
It is interesting to mention that from a technical point of view it is
mostly more simple to make first the integration over the variable $q$
in this formula for simple cases of Radon transforms such as, for example,
for squeezed coherent states whereas the direct integration over the angle
of the Radon transform multiplied by phase factors is connected with
considerable technical difficulties.

   It arises the question whether solutions of Eq.(5.4) with unequal 
divisions of the unit circle are possible or not. Such solutions are, in 
principle, possible if one uses $n+1$ inequivalent discrete angles 
( pairs of angles $\varphi$ and $\varphi + \pi$ are equivalent ). However, 
it seems to be difficult to find general solutions for arbitrary $n$ in 
explicit form. For low orders of $n$ it is possible to solve the system of 
algebraic equations obtained for discrete angles. In such a way one finds 
for the first-order moments
\begin{eqnarray}
\langle a \varrho \rangle \!\!&=& \!\!
\frac{1}{2\sqrt{2}}\int_{-\infty}^{+\infty}dq\,\left\{
\frac{\exp(i\varphi_{1}) \breve{W}(\cos\varphi_{0},\sin\varphi_{0};q)}  
{i\sin(\varphi_{1}-\varphi_{0})}+
\frac{\exp(i\varphi_{0}) \breve{W}(\cos\varphi_{1},\sin\varphi_{1};q)}  
{i\sin(\varphi_{0}-\varphi_{1})}\right\} \nonumber\\
&&\!\! \times H_1\left(\frac{q}{\sqrt{\hbar}}\right), 
\qquad \langle a^\dagger \varrho \rangle = \langle a \varrho \rangle^*,
\qquad H_1(x)=2x,
\end{eqnarray}
and for the second-order moments
\begin{eqnarray}
\langle a^2 \varrho \rangle \!\!&=&\!\!
\frac{1}{8}\int_{-\infty}^{+\infty}dq\,\Bigg\{ 
\frac{\exp\left(i(\varphi_{1}+\varphi_{2})\right)
\breve{W}(\cos\varphi_{0},\sin\varphi_{0};q)}  
{\sin(\varphi_{1}-\varphi_{0})\sin(\varphi_{0}-\varphi_{2})}
\nonumber\\&&\!\!+
\frac{\exp\left(i(\varphi_{2}+\varphi_{0})\right)
\breve{W}(\cos\varphi_{1},\sin\varphi_{1};q)}  
{\sin(\varphi_{1}-\varphi_{0})\sin(\varphi_{2}-\varphi_{1})}
\nonumber\\&&\!\!+
\frac{\exp\left(i(\varphi_{0}+\varphi_{1})\right)
\breve{W}(\cos\varphi_{2},\sin\varphi_{2};q)}  
{\sin(\varphi_{2}-\varphi_{1})\sin(\varphi_{0}-\varphi_{2})}\Bigg\}
H_2\left(\frac{q}{\sqrt{\hbar}}\right),\nonumber\\
\langle a^\dagger a \varrho \rangle\!\!&=&\!\!
-\frac{1}{8}\int_{-\infty}^{+\infty}dq\,\Bigg\{ 
\frac{\cos\left(\varphi_{2}-\varphi_{1}\right)
\breve{W}(\cos\varphi_{0},\sin\varphi_{0};q)}  
{\sin(\varphi_{1}-\varphi_{0})\sin(\varphi_{0}-\varphi_{2})}
\nonumber\\&&\!\!+
\frac{\cos\left(\varphi_{0}-\varphi_{2}\right)
\breve{W}(\cos\varphi_{1},\sin\varphi_{1};q)}  
{\sin(\varphi_{1}-\varphi_{0})\sin(\varphi_{2}-\varphi_{1})}
\nonumber\\&&\!\!+
\frac{\cos\left(\varphi_{1}-\varphi_{0}\right)
\breve{W}(\cos\varphi_{2},\sin\varphi_{2};q)}  
{\sin(\varphi_{2}-\varphi_{1})\sin(\varphi_{0}-\varphi_{2})}\Bigg\}
H_2\left(\frac{q}{\sqrt{\hbar}}\right), 
\nonumber\\[2mm] \langle a^{\dagger\,2} \varrho \rangle\ \!\!&=&\!\! 
\langle a^2 \varrho \rangle^*, \qquad H_{2}(x)=4x^2-2.
\end{eqnarray}
It is easy to specify these solutions. For example, for $\varphi_0=0,
\varphi_1=\pi/4, \varphi_2=\pi/2$ it follows from (5.10)
\begin{eqnarray}
\langle a^2 \varrho \rangle \!\!&=&\!\!
\frac{1}{8}\int_{-\infty}^{+\infty}dq\,\Bigg\{ 
(1-i)\breve{W}(1,0;q)
+i2 \breve{W}\left(\frac{\sqrt{2}}{2},\frac{\sqrt{2}}{2};q\right)-
(1+i)\breve{W}(0,1;q)\Bigg\}H_2\left(\frac{q}{\sqrt{\hbar}}\right),
\nonumber\\
\langle a^\dagger a \varrho \rangle\!\!&=&\!\!
\frac{1}{8}\int_{-\infty}^{+\infty}dq\,\Bigg\{   
\breve{W}(1,0;q)+\breve{W}(0,1;q)\Bigg\}
H_2\left(\frac{q}{\sqrt{\hbar}}\right),
\end{eqnarray}
and for harmonic division $\varphi_0=0, \varphi_1=\pi/3, \varphi_2=2\pi/3$
\newpage
\begin{eqnarray}
\langle a^2 \varrho \rangle \!\!&=&\!\!
\frac{1}{6}\int_{-\infty}^{+\infty}dq\,\Bigg\{ 
\breve{W}(1,0;q)-
\frac{1-i\sqrt{3}}{2}\breve{W}\left(\frac{1}{2},\frac{\sqrt{3}}{2};q\right)-
\frac{1+i\sqrt{3}}{2}\breve{W}\left(-\frac{1}{2},\frac{\sqrt{3}}{2};q\right)
\Bigg\}\nonumber\\ &&\!\!\times
H_2\left(\frac{q}{\sqrt{\hbar}}\right),
\nonumber\\
\langle a^\dagger a \varrho \rangle\!\!&=&\!\!
\frac{1}{12}\int_{-\infty}^{+\infty}dq\,\Bigg\{   
\breve{W}(1,0;q)+
\breve{W}\left(\frac{1}{2},\frac{\sqrt{3}}{2};q\right) +
\breve{W}\left(-\frac{1}{2},\frac{\sqrt{3}}{2};q\right)\Bigg\}
H_2\left(\frac{q}{\sqrt{\hbar}}\right).
\end{eqnarray}
We are not sure to what extent Eqs.(5.9) and (5.10) can be  
explicitly generalized. The harmonic division has the disadvantage that one 
must change the division when going from moments of the $n$--th order to 
moments of $(n+1)$--th order. If one is only interested up to moments of 
the 4--th order then one can take the harmonic division in 5 angles according 
to $\varphi_0+(m\pi)/5$ with $m=0,1,\ldots,4$ and can use for the lower 
moments subsets of these angles corresponding to anharmonic divisions. In 
photon statistics one considers only moments $\langle N^l \varrho \rangle$ 
corresponding to linear combinations of normally ordered moments $\langle
a^{\dagger\,k} a^k \varrho \rangle$ up to the order $l$. It is quite possible 
that some other special solutions of the systems of equations (5.4) 
corresponding to anharmonic divisions can be found.

\setcounter{chapter}{6}
\setcounter{equation}{0}
\chapter*{6. Alternative method of the reconstruction of the matrix 
elements} 

   The matrix elements of the density operator in Fock-state representation
can be also reconstructed from the normally ordered moments of the density
operator. If we use the reconstruction of the normally ordered moments
from the Radon transform that was discussed in the last section as an 
intermediate step to the reconstruction of the matrix elements then we arrive 
at a new relatively simple alternative representation of the pattern functions. 
We now consider the derivation.

   From the representation of the density operator by the normally ordered
moments in Eq.(4.1) one obtains by forming the matrix elements with  Fock
states
\begin{eqnarray}
\langle m|\varrho|n \rangle  \!\!&=&\!\! \sum_{k=0}^{\infty} 
\sum_{l=0}^{\infty} \langle m|a_{k,l}|n\rangle \langle a^{\dagger\,k}a^l
\varrho \rangle \nonumber\\ &=&\!\!  \sum_{k=0}^{\infty} \sum_{l=0}^{\infty}
\sum_{j=0}^{\{k,l\}}
\frac{(-1)^j \langle m|l-j \rangle \langle k-j|n \rangle}{j!\sqrt{(k-j)!
(l-j)!}}\langle a^{\dagger\,k}a^l\varrho \rangle \nonumber \\ &=&\!\!  
\frac{1}{\sqrt{m!n!}} \sum_{j=0}^{\infty} \frac{(-1)^j}{j!} 
\langle a^{\dagger\,n+j}a^{m+j}\varrho \rangle.
\end{eqnarray}
By inserting the reconstruction formula for the normally ordered moments 
from the Radon transform in the form averaged over the angle (5.8) one finds
after changing the order of summation and integration
\begin{eqnarray}
\langle m|\varrho|n \rangle \!\!&=&\!\! \frac{1}{\pi} \int_{0}^{\pi} 
d\varphi\,\exp\left\{i(m-n)\varphi \right\} \int_{-\infty}^{+\infty} dq\,
\breve{W}(\cos\varphi,\sin\varphi;q) \nonumber\\ &&\!\! \times
\frac{1}{\sqrt{2^{m+n}m!n!}} 
\sum_{j=0}^{\infty} \frac{(m+j)!(n+j)!}{j!(m+n+2j)!}\left(-\frac{1}{2}
\right)^j H_{m+n+2j}\left(\frac{q}{\sqrt{\hbar}}\right) \nonumber\\
&\equiv&\!\! \frac{1}{\pi} \int_{0}^{\pi} 
d\varphi\,\exp\left\{i(m-n)\varphi \right\} \int_{-\infty}^{+\infty} dq\,
\breve{W}(\cos\varphi,\sin\varphi;q)\,F'_{m,n}\left(\frac{q}{\sqrt{\hbar}}\right).
\end{eqnarray}
This means that we derived a representation of the pattern functions
$F'_{m,n}(x)$ by an infinite series over Hermite polynomials of even or
odd order in increasing steps of $2$ in the indices with some coefficients 
and in dependence on the even or odd order of $m+n$ 
\begin{eqnarray}
F'_{m,n}(x)\!\!&=&\!\!\frac{1}{\sqrt{2^{m+n}m!n!}} 
\sum_{j=0}^{\infty} \frac{(m+j)!(n+j)!}{j!(m+n+2j)!}\left(-\frac{1}{2}
\right)^j H_{m+n+2j}(x)
\nonumber\\&=&\!\! 
\frac{\partial}{\partial x} \left\{
\frac{1}{2\sqrt{2^{m+n}m!n!}} 
\sum_{j=0}^{\infty} \frac{(m+j)!(n+j)!}{j!(m+n+2j+1)!}\left(-\frac{1}{2}
\right)^j H_{m+n+2j+1}(x)\right\}. 
\end{eqnarray}
The second representation in (6.3) by a derivative of a function with 
respect to $q$ was made because it corresponds to the general structure
of the pattern functions as derived in 
[18--20]
where the content
in braces can be represented as a product of the only normalizable 
eigenfunction of the number operator with a nonnormalizable eigenfunction 
of this operator or linear combinations of such products  
( superpositions of the normalizable with each nonnormalizable eigenfunction 
give in every case nonnormalizable eigenfunctions, however, with 
no determined parity, in general ). 
We wrote $F'_{m,n}(x)$ for the pattern functions because they must not    
necessarily be identical with the pattern functions $F_{m,n}(x)$ derived 
in Eq.(4.27). We now investigate this new phenomenon of the nonuniqueness 
of the pattern functions.
   
   In the last section it was found in Eq.(5.4) that the multiplication of the
Radon transform with a Hermite polynomial $H_{s}\left(q/\sqrt{\hbar}\right)$
and its integration over $q$ leads to a linear combination of $(s+1)$ 
normally ordered moments of degree $s$ multiplied by some binomials 
coefficients and phase factors ${\rm e}^{i(2k-s)\varphi}$ with 
$k=0,1,\ldots,s$. In the reconstruction formula of the Fock-state matrix 
elements from the Radon transform of the Wigner quasiprobability (4.24) the 
pattern functions $F_{m,n}(x)$ have to be multiplied by phase factors 
${\rm e}^{i(m-n)\varphi}$ and integrated over $\varphi$ from $0$ to $\pi$.
Because of
\begin{equation}
\frac{1}{\pi}\int_{0}^{\pi} d\varphi\,\exp(i2l\varphi)=\delta_{l,0},
\end{equation}
two pattern functions $F_{m,n}(x)$ and $F'_{m,n}(x)$ connected by   
\begin{equation}
F'_{m,n}(x)=F_{m,n}(x)+\sum_{k=1}^{\left[\left|\frac{m-n}{2}\right|\right]}
c_{k}H_{|m-n|-2k}(x),
\end{equation}
where $[\nu]$ denotes the integer part of $\nu$ and 
where $c_{k}$ are arbitrary coefficients, lead to the same result for the  
reconstruction of the Fock-state matrix elements $\langle m|\varrho|n 
\rangle$ and are equivalent. It is not proved that (6.5) is the most general 
form of the nonuniqueness of the pattern functions but it seems so and it is 
the only form of the nonuniqueness which plays a role in the present paper.
The pattern functions obtained in (4.27) become identical with
the pattern functions in (6.3) if we only formally extend the summation to 
all possible negative summation indices $j$, i.e.
\begin{eqnarray}   
F_{m,n}(x)\!\!&=&\!\!\frac{1}{\sqrt{2^{m+n}m!n!}} 
\sum_{j=-\left[\frac{m+n}{2}\right]}^{\infty} 
\frac{(m+j)!(n+j)!}{j!(m+n+2j)!}\left(-\frac{1}{2}\right)^j H_{m+n+2j}(x)
\nonumber\\ &=& \!\! F'_{m,n}(x) \nonumber\\ &&\!\!
+\frac{1}{\sqrt{2^{|m-n|}m!n!}}
\sum_{k=1}^{\left[\left|\frac{m-n}{2}\right|\right]} 
\frac{\left(k-1+\{m,n\}\right)!\left(|m-n|-k\right)!}
{(k-1)!\left(|m-n|-2k\right)!}(-2)^k H_{|m-n|-2k}(x),
\nonumber\\ 
\end{eqnarray}
where $\{m,n\} \equiv {\rm Min}(m,n)$. 
All investigated special cases are in agreement with this statement but the 
general proof is not made up to now. In special cases one finds 
\begin{eqnarray}
\!\!\!\!\!&&F_{n,n}(x)=F'_{n,n}(x),\quad F_{n+1,n}(x)=F'_{n+1,n}(x),
\nonumber\\ \!\!\!\!\!&& F_{n+2,n}(x)=F'_{n+2,n}(x)-\sqrt{\frac{n!}{(n+2)!}}
H_{0}(x), \quad F_{n+3,n}(x)=F'_{n+3,n}(x)-\sqrt{\frac{2n!}{(n+3)!}}H_{1}(x).
\nonumber\\
\end{eqnarray}

   Let us discuss some properties of the pattern functions $F'_{m,n}(x)$.
First of all, they are symmetric in the indices  and possess the parity 
$(-1)^{m+n}$ as $F_{m,n}(x)$ too
\begin{equation}
F'_{m,n}(x)=F'_{n,m}(x),\quad F'_{m,n}(-x)=(-1)^{m+n}F'_{m,n}(x).
\end{equation}
The summations at the point $x=0$ can be exactly accomplished for not very 
large even differences $|m-n|$ whereas for all odd differences the 
corresponding Hermite polynomials in (6.3) vanish for $x=0$. One obtains
\cite{ulf}
\begin{eqnarray}
&& F'_{n,n}(0)=(-1)^n 2,\quad F'_{n+2,n}(0)=(-1)^{n+1}\frac{2n+3}{\sqrt{
(n+2)(n+1)}},\quad
F'_{n+2l+1,n}(0)=0.
\end{eqnarray}
The first maxima and minima of $F'_{n,n}(x)=F_{n,n}(x)$ for $x\neq0$ come 
for increasing $n$ very near to the values $\pm 2$ but are not exactly equal 
to these values \cite{ulf}. Some similarities in the form of the pattern 
functions to the corresponding products of wave functions 
$\langle q|m \rangle \langle n| q \rangle$ can be explained by the following 
expansion of products of Hermite polynomials which can be proved by complete 
induction ( see Eq.10.13 (37) ) in \cite{bat} )
\begin{eqnarray}
& &\langle q|m \rangle \langle n| q \rangle = \frac{1}{\sqrt{\pi\hbar}}
\exp\left(-\frac{q^2}{\hbar}\right) \frac{1}{\sqrt{2^{m+n}m!n!}}
H_{m}\left(\frac{q}{\sqrt{\hbar}}\right) H_{n}\left(\frac{q}{\sqrt{\hbar}}
\right),\nonumber\\
& & H_{m}(x)H_{n}(x)=\sum_{j=0}^{\{m,n\}}\frac{m!n!}{j!(m-j)!(n-j)!}2^j
H_{m+n-2j}(x).
\end{eqnarray}
The first polynomial terms in the expansions in (6.3) and (6.10) are both
$H_{m+n}(x)$. Contrary to (6.3), the next terms in the expansion in (6.10)
are proportional to Hermite polynomials with indices decreasing in steps of
2 but if the normalized Gaussian function $ \exp(-x^2)/\sqrt{\pi}$ is 
included into the expansion in terms of Hermite polynomials, then we get 
a more complicated formula which we do not derive here. Thus one has 
similarities in the representation of the products of wave functions and
corresponding pattern functions by expansions in Hermite polynomials.

   The representation (6.3) of the pattern functions is appropriate for
their calculation by a computer with good accuracy for not too large 
arguments ( say $|x|< 5$ with $\sim 50$ initial sum terms and not too high 
$(m,n)$, e.g., for $m=n=0$ one obtains by 15 initial terms an approximation 
where the deviations begin for $|x|\approx 5$ ). For large values of the 
argument one can use the highest power in the Hermite polynomials $H_{n}(x)$ 
that means $(2x)^n$ as an approximation. 
It is interesting that the Hermite polynomials $H_{n}(x)$ can be obtained 
from these asymptotic functions $(2x)^n$ by the following convolution  
\cite{fan,w5}
\begin{equation}
H_{n}(x)=\exp\left(-\frac{1}{4}\frac{\partial^2}{\partial x^2}\right)(2x)^n    
=\frac{1}{\sqrt{-\pi}}\exp\left(x^2\right)*(2x)^n.
\end{equation}
One can prove by Taylor series expansion of the exponential function
that one obtains by this convolution the explicit representation of the 
Hermite polynomials. By applying this to Eq.(6.3) one finds
\begin{equation}
F'_{m,n}(x) =
\exp\left(-\frac{1}{4}\frac{\partial^2}{\partial x^2}\right) \frac{
(\sqrt{2}\,x)^{m+n}}{\sqrt{m!n!}}\sum_{j=0}^{\infty}\frac{(m+j)!(n+j)!}
{j!(m+n+2j)!}\left(-2x^2\right)^j.
\end{equation}
If we omit the convolution operator $\exp\left(-(1/4) \partial^2/\partial x^2
\right)$ in this relation we get an asymptotic representation of the pattern 
functions for large values $|x|$. 

   If we use the explicit representation of the Hermite polynomials in the
pattern function $F'_{0,0}(x)=F_{0,0}(x)$ we arrive, after reordering of the 
sum terms and accomplishing one of the sums, at the following Taylor series 
representation
\begin{eqnarray}
F_{0,0}(x)\!\!&=&\!\!\sum_{j=0}^{\infty}\frac{j!}{(2j)!}\left(-\frac{1}{2}
\right)^j H_{2j}(x) \nonumber\\ &=&\!\! \frac{\partial}{\partial x}\frac{1}{2}
\sum_{j=0}^{\infty}\frac{j!}{(2j+1)!}\left(-\frac{1}{2}\right)^j H_{2j+1}(x)
\nonumber\\ &=&\!\! 2\sum_{k=0}^{\infty}\frac{(-1)^k k!}{(2k)!}(2x)^{2k}.
\end{eqnarray}
This can be also represented in the forms
\begin{eqnarray}
F_{0,0}(x)\!\!&=&\!\! \frac{\partial}{\partial x} \left\{2\exp\left(-x^2
\right)\int_{0}^{x}dt \,\exp\left(t^2\right)\right\}\nonumber\\ &=&\!\! 
\frac{\partial}{\partial x}\left\{ \sqrt{2} \exp\left(
-\frac{x^2}{2}\right)\frac{i}{2}\left\{D_{-1}\Big(i\sqrt{2}x\Big)-
D_{-1}\Big(-i\sqrt{2}x\Big)\right\}\right\} \nonumber\\&=&\!\! 
\exp\left(-\frac{x^2}{2}\right) \left\{D_{-2}\Big(i\sqrt{2}x\Big)+ 
D_{-2}\Big(-i\sqrt{2}x\Big)\right\}.
\end{eqnarray}
where $D_{\nu}(z)$ denotes the functions of the parabolic cylinder. Let us 
show that the content in braces in Eq.(6.3) is proportional to the product 
of the normalizable eigenfunction of order $m$ and the nonnormalizable 
eigenfunction of order $n$. For this purpose we make the following 
transformation
\begin{eqnarray}
&&\!\! \left\{\left(x-\frac{\partial}{\partial x}\right)^m\exp\left(
-\frac{x^2}{2}\right)\right\}\left\{ \left(x-\frac{\partial}{\partial x}
\right)^n 2\exp\left(-\frac{x^2}{2}\right)\int_{0}^{x}du\,\exp\left(u^2\right)
\right\} \nonumber\\ &=& \!\! 2 H_{m}(x) \left(-\frac{\partial}{\partial x}
\right)^n \left\{\exp\left(-x^2\right)\int_{0}^{x}du\,\exp\left(u^2\right) 
\right\}
\nonumber\\ &=&\!\! \frac{1}{2} H_{m}(x) \left(-\frac{\partial}{\partial x} 
\right)^n \sum_{k=0}^{\infty} \frac{k!}{(2k+1)!}\left(-\frac{1}{2}\right)^k
H_{2k+1}(x) \nonumber\\ &=& \!\! \frac{1}{2} 
\sum_{j=-\left[\frac{m+n+1}{2}\right]}^{\infty}
\frac{(m+j)!(n+j)!}{j!(m+n+2j+1)!}\left(-\frac{1}{2}\right)^j H_{m+n+2j+1}(x).
\end{eqnarray}
In the first step we brought the operator $(x-\partial/\partial x)^n$ from
the right to the left of the function $\exp(-x^2/2)$. Then we applied (6.13)
and substituted the product of Hermite polynomials by sums over Hermite
polynomials according to a known formula ( \cite{bat}, 10.13 (36) ) which
can be proved by complete induction. After rearrangement of the double
sum one sum is calculated by using
\begin{equation}
\sum_{l=0}^m\frac{(-1)^l m!(n+j+l)!}{l!(m-l)!(n+2j+1+l)!}=\frac{(m+j)!(n+j)!}
{j!(m+n+2j+1)!},
\end{equation}
which can be proved again by complete induction ( $m\rightarrow m+1$ because
it is true for $m=0$ with arbitrary $n$ and $j$ ). Thus there is directly 
proved the connection between two essentially different representations of 
the pattern functions. We think that the representation of the pattern
functions by infinite sums over Hermite polynomials possesses some 
advantages in comparison to the representation by sums over functions
of the parabolic cylinder of imaginary argument. 

\setcounter{chapter}{7}
\setcounter{equation}{0}
\chapter*{7. Differential equations for products of Hermite functions 
and orthogonality relations} 

Initiated by Richter \cite{ri1} to explain the structure of the pattern
functions as first derivatives of the product of the normalizable and one
nonnormalizable eigenfunction of the number operator in "position" 
representation it was found \cite{riw,riwu2} that the most appropriate 
approach to solve this problem is to consider the differential equation for 
products of Hermite functions and to establish the orthogonality relations 
with the help of the adjoint differential equation. This will be presented 
here. The problem solved in \cite{riw} was from one point of view more 
general because there were considered products of eigenfunctions of more 
general Hamilton operators for one degree of freedom consisting of the 
usual part from kinetic energy and of a part from arbitrary potential energy 
but from the other point of view it was more special because there were 
considered only products of eigenfunctions to the same eigenvalue. This last 
restriction is omitted in \cite{riwu2}. The most interesting case for the 
nondiagonal elements corresponding to products of eigenfunctions to different 
eigenvalues is, however, the case of the harmonic oscillator because for the 
most other interesting one-dimensional potentials defined over the whole 
coordinate axis $R$ one has no degeneracy of energy differences in the 
eigenvalue spectrum. The multiplication of the Radon transform with phase 
factors and integration over the angle in case of the harmonic oscillator 
sorts out the Fock-state matrix elements corresponding to equal energy 
differences of the energy eigenvalue spectrum.

The differential equations for both series of normalizable and 
nonnormalizable excitation states $h_n(x)$ and $g_n(x)$ of the harmonic 
oscillator in position representation are
( as before we use the abbreviation $x\equiv q/\sqrt{\hbar}$ ) 
\begin{equation}
\left\{\frac{\partial^2}{\partial x^2}-x^2+2n+1\right\}f_n(x)=0, \quad 
\longleftrightarrow \quad \left(\frac{x^2}{2} -\frac{1}{2}
\frac{\partial^2}{\partial x^2}\right)f_n(x)=\left(n+\frac{1}{2}\right)f_n(x),
\end{equation}
where $f_n(x)$ is written for an arbitrary of the functions $h_n(x)$ or 
$g_n(x)$ or their linear combinations. The normalizable functions $h_n(x)$ 
are the Hermite functions. They possess the parity $(-1)^n$ and are given up 
to complex normalization factor in a unique way and are connected with the 
Hermite polynomials $H_n(z)$ and the functions of the parabolic cylinder 
$D_n(z)$ as follows
\begin{eqnarray}
h_n(x)\!\!&=&\!\!\frac{1}{\pi^{\frac{1}{4}}}\exp\left(-\frac{x^2}{2}\right)
\frac{H_n(x)}{\sqrt{2^n n!}}\nonumber\\ &=& \!\! 
\frac{D_n\left(\sqrt{2}x\right)}{\pi^{\frac{1}{4}}\sqrt{n!}} \nonumber\\ 
&=&\!\! \frac{\sqrt{2\,n!}}{\pi^{\frac{3}{4}}}\,\frac{i^n}{2}\left\{D_{-1-n}
\Big(i\sqrt{2}x\Big)+(-1)^{n}D_{-1-n}\Big(-i\sqrt{2}x\Big)\right\}.
\end{eqnarray}
They satisfy the following orthonormality and completeness relations
\begin{equation}
\int_{-\infty}^{+\infty}dx\,h_m(x)h_n(x)=\delta_{m,n},\qquad 
\sum_{n=0}^{\infty}h_n(x)h_n(y)=\delta(x-y).
\end{equation}
All other linearly independent solutions of equation (7.1) are 
nonnormalizable but among them one can choose $g_{n}(x)$ in a unique way as
the solutions with parity $(-1)^{n+1}$ defined in (4.29). The explicit form
of these solutions expressed by the functions of the parabolic cylinder is
given by
\begin{eqnarray}
g_{n}(x)\!\!&=&\!\!\pi^{\frac{1}{4}}\sqrt{2\,n!}\,\frac{i^{n+1}}{2}
\left\{D_{-1-n}\Big(i\sqrt{2}x\Big)-(-1)^{n}D_{-1-n}\Big(-i\sqrt{2}x\Big)
\right\}\nonumber\\&=&\!\!\frac{2\pi^{\frac{1}{4}}}{\sqrt{2^n n!}}
\exp\left(\frac{x^2}{2}\right) \bigg\{H_{n}(x)F(x)-\sum_{k=0}^{n-1}
\frac{n!}{k!(n-k)!}i^{n-1-k}H_{k}(x)H_{n-1-k}(ix)\bigg\},
\end{eqnarray}
where $F(x)$ denotes the Dawson integral defined in (4.18) ( see also (4.20)
for the connection to the error function of imaginary argument ). From (7.2)
and (7.4) one finds that functions proportional to $D_{-1-n}\left(i\sqrt{2}x
\right)$ and $D_{-1-n}\left(-i\sqrt{2}x\right)$ could be also chosen as two
linearly independent solutions of the equations (7.1) but they are both 
nonnormalizable and do not possess definite parity.

Let us now consider the differential equation for the product $f_m(x)g_n(x)$
of two functions $f_m(x)$ and $g_n(x)$ both satisfying differential 
equations of the form (7.1) ( $g_n(x)$ is here not necessarily the special
eigenfunction with parity $(-1)^{n+1}$ ). Then one finds up to second-order 
derivatives 
\begin{eqnarray}
\frac{\partial}{\partial x}\left\{f_m(x)g_n(x)\right\} \!\!&=&\!\!  
f_m^{(1)}(x)g_n(x)+f_m(x)g_n^{(1)}(x),\nonumber\\ 
\frac{\partial^2}{\partial x^2}\left\{f_m(x)g_n(x)\right\} \!\!&=&\!\!
f_m^{(2)}(x)g_n(x)+2f_m^{(1)}(x)g_n^{(1)}(x)+f_n(x)g_n^{(2)}(x) 
\nonumber\\&=&\!\! 2f_m^{(1)}(x)g_n^{(1)}(x) +2\Big(x^2-(m+n+1)\Big)
f_m(x)g_n(x).
\end{eqnarray}
In the next step by further differentiation and by applying (7.1) one finds
\begin{eqnarray}
&&\!\!\left\{\frac{\partial^3}{\partial x^3}
-2\left(\Big(x^2-(m+n+1)\Big)\frac{\partial}{\partial x} 
+\frac{\partial}{\partial x}\Big(x^2-(m+n+1)\Big)\right)
\right\} f_m(x)g_n(x) \nonumber\\
&=&\!\! -2(m-n)\left\{f_m(x)g_n^{\,(1)}(x)-f_m^{(1)}(x)g_n(x)\right\},
\end{eqnarray}
written in a symmetrical form with respect to the derivatives.
It is a pure differential equation of third order for $f_m(x)g_n(x)$ only 
in case of $m=n$. This is understandable because in this case we can form 
3 linearly independent combinations of the basic solutions, for example, 
$(h_{n}(x))^2, h_{n}(x)g_{n}(x), (g_{n}(x))^2$ if $h_{n}(x)$ and $g_{n}(x)$
are two linearly independent solutions of (7.1).
For $m \neq n$ one needs a further differentiation of this equation to
obtain the following fourth-order differential equation 
\begin{eqnarray}
\!\!&&\!\!\left\{\frac{\partial^4}{\partial x^4}
-2\frac{\partial}{\partial x}\left(\Big(x^2-(m+n+1)\Big)\frac{\partial}
{\partial x}+ \frac{\partial}{\partial x}\Big(x^2-(m+n+1)\Big)\right)
+4(m-n)^2 \right\} \nonumber\\ \!\!& &\!\! \times
f_m(x)g_n(x)= 0.
\end{eqnarray}
We have here 4 linearly independent solutions of this differential equation
as which can be chosen, for example, the products $h_{m}(x)h_{n}(x), 
h_{m}(x)g_{n}(x), g_{m}(x)h_{n}(x), g_{m}(x)g_{n}(x)$. The operator in  
the differential equation (7.7) is neither selfadjoint nor anti-selfadjoint.
The adjoint differential equation to (7.7) for functions $X_{m,n}(x)$ is 
\begin{eqnarray}
\!\!&&\!\!\left\{\frac{\partial^4}{\partial x^4}
-2\left(\Big(x^2-(m+n+1)\Big)\frac{\partial}{\partial x}+ \frac{\partial}
{\partial x}\Big(x^2-(m+n+1)\Big)\right)\frac{\partial}{\partial x}
+4(m-n)^2 \right\} \nonumber\\ \!\!& &\!\! \times
X_{m,n}(x)= 0.
\end{eqnarray}
If we differentiate this equation once more then we get immediately
\begin{eqnarray}
\!\!&&\!\!\left\{\frac{\partial^4}{\partial x^4}
-2\frac{\partial}{\partial x}\left(\Big(x^2-(m+n+1)\Big)\frac{\partial}
{\partial x}+ \frac{\partial}{\partial x}\Big(x^2-(m+n+1)\Big)\right)
+4(m-n)^2 \right\} \nonumber\\ \!\!& &\!\! \times
\frac{\partial}{\partial x}X_{m,n}(x)= 0.
\end{eqnarray}
This means that $(\partial/\partial x)X_{m,n}(x)$ satisfies the 
differential equation (7.7) for products of functions $f_{m}(x)g_{n}(x)$
and their linear combinations.

We now make the proper derivation of the orthogonality relations. For
this purpose we first write down the following two differential equations 
according to (7.7) and (7.8)
\begin{eqnarray}
\!\!&&\!\!\left\{\frac{\partial^4}{\partial x^4}
-2\frac{\partial}{\partial x}\left(\Big(x^2-(m+n+1)\Big)\frac{\partial}
{\partial x}+ \frac{\partial}{\partial x}\Big(x^2-(m+n+1)\Big)\right)
+4(m-n)^2 \right\} \nonumber\\ \!\!& &\!\!  \times Y_{m,n}(x)= 0,
\nonumber\\[2mm] \!\!&&\!\!\left\{\frac{\partial^4}{\partial x^4}
-2\left(\Big(x^2-(k+l+1)\Big)\frac{\partial}{\partial x}+ \frac{\partial}
{\partial x}\Big(x^2-(k+l+1)\Big)\right)\frac{\partial}{\partial x}
+4(k-l)^2 \right\} \nonumber\\ \!\!& &\!\! \times
X_{k,l}(x)= 0,
\end{eqnarray}
where $Y_{m,n}(x)$ is written for arbitrary products $f_{m}(x)g_{n}(x)$
of solutions of (7.1) or their linear combinations.
If we multiply the first equation with $X_{k,l}(x)$ and the second equation 
with $Y_{m,n}(x)$ and subtract the obtained equations then we find the 
following possible representation of the resulting equation
\begin{eqnarray}
0\!\!\!&=&\!\!\! \frac{\partial}{\partial x}\Bigg\{
X_{k,l}(x) Y_{m,n}^{(3)}(x)-X_{k,l}^{(1)}(x)Y_{m,n}^{(2)}(x)
+X_{k,l}^{(2)}(x)Y_{m,n}^{(1)}(x)-X_{k,l}^{(3)}(x)Y_{m,n}(x)   
\nonumber\\ &&\!\!\!
-2\Big(2x^2-(m+n+k+l+2)\Big)\Big\{X_{k,l}(x)Y_{m,n}^{(1)}(x)
-X_{k,l}^{(1)}(x)Y_{m,n}(x)\Big\}-4x X_{k,l}(x)Y_{m,n}(x)\Bigg\}
\nonumber\\ &&\!\!\!
+2(m+n-k-l)\left\{X_{k,l}(x)Y_{m,n}^{(2)}(x)+X_{k,l}^{(2)}(x)Y_{m,n}(x)
\right\} \nonumber\\[2mm] && \!\!\!
+4\left((m-n)^2-(k-l)^2\right)X_{k,l}(x)Y_{m,n}(x).
\end{eqnarray}
By integration of this equation over the whole coordinate axis $R$ under
the assumption that the products $X_{k,l}^{(r)}(x)Y_{m,n}^{(s)}(x)$ vanish
for $x\,\rightarrow\,\pm \infty$ one finds 
\begin{eqnarray}
&&\!\!(m+n-k-l)
\int_{-\infty}^{+\infty}dx\,\left\{X_{k,l}(x)Y_{m,n}^{(2)}(x)
+X_{k,l}^{(2)}(x)Y_{m,n}(x) \right\}
\nonumber\\ && \!\!+2\left\{(m-n)^2-(k-l)^2\right\}
\int_{-\infty}^{+\infty}dx\,X_{k,l}(x)Y_{m,n}(x) = 0.
\end{eqnarray}
The first of the integrals in (7.12) can be transformed by partial integration
as follows
\begin{eqnarray}
&&\!\!\int_{-\infty}^{+\infty}dx\,\left\{X_{k,l}(x)Y_{m,n}^{(2)}(x)
+X_{k,l}^{(2)}(x)Y_{m,n}(x) \right\}\nonumber\\ &=&\!\!
\int_{-\infty}^{+\infty}dx\,\left\{
-X_{k,l}^{(1)}(x)\left(\frac{\partial}{\partial x}Y_{m,n}(x)\right)
+\left(\frac{\partial}{\partial x}X_{k,l}^{(1)}(x)\right)Y_{m,n}(x) \right\}    
\nonumber\\ &=&\!\! 2 \int_{-\infty}^{+\infty}dx\,
\left(\frac{\partial}{\partial x}X_{k,l}^{(1)}(x)\right)Y_{m,n}(x).
\end{eqnarray}
From (7.13) and (7.12) one can derive different orthogonality relations. 
Consider the case 
\begin{equation}
m-n=k-l \equiv -j,\:\longrightarrow\: l=k+j,\quad n=m+j, \quad m+n-k-l=2(m-k).
\end{equation}
Since the integral becomes nonvanishing only for $m-k=0$ one obtains
\begin{equation}
\int_{-\infty}^{+\infty}dx\,\left(\frac{\partial}{\partial x}
X_{k,k+j}^{(1)}(x)\right)Y_{m,m+j}(x) = \delta_{k,m}
\int_{-\infty}^{+\infty}dx\,
\left(\frac{\partial}{\partial x}X_{m,m+j}^{(1)}(x)\right)Y_{m,m+j}(x).
\end{equation}
In particular, by choosing 
\begin{equation}
X_{k,l}^{(1)}(x)= h_{k}(x)g_{l}(x), \quad Y_{m,n}(x)=h_{m}(x)h_{n}(x),
\end{equation}
where $g_{l}(x)$ denotes the nonnormalizable solution (7.4) of Eq.(7.1) with 
parity $(-1)^{n+1}$ one finds from (7.15) with the help of (7.13)
\begin{eqnarray}
&& \!\!\int_{-\infty}^{+\infty}dx\,\left(\frac{\partial}{\partial x}
h_{k}(x)g_{k+j}(x)\right)h_{m}(x)h_{m+j}(x) \nonumber\\ &=& \!\!
\delta_{k,m} \frac{1}{2} \int_{-\infty}^{+\infty}dx\,\Bigg\{
\left(\frac{\partial}{\partial x}h_{m}(x)g_{m+j}(x)\right)h_{m}(x)h_{m+j}(x)
\nonumber\\ && \!\!
-h_{m}(x)g_{m+j}(x)\left(\frac{\partial}{\partial x}h_{m}(x)h_{m+j}(x)\right)
\Bigg\} \nonumber\\ &=&\!\! \delta_{k,m}\frac{1}{2}W(h_{m+j}(x),g_{m+j}(x)) 
\int_{-\infty}^{+\infty}dx\,h_{m}(x)h_{m}(x) \nonumber\\[2mm] &=&\!\! 
\delta_{k,m},
\end{eqnarray}
where the special value of the Wronskian given in (4.29) was used.
The special case $j=0$ of the 
orthogonality relations (7.17) can be obtained from the third-order 
differential equation (7.6) for $m=n$ in an easier way but for shortness 
we did not separately give its derivation. 

Using the representation of $h_m(x)$ by Hermite functions given in (7.2)
and the representation of $g_n(x)$ by functions of the parabolic cylinder 
given in (7.4) one obtains by differentiation of their product the following 
representation of the pattern functions (4.25)
\begin{eqnarray}
F''_{m,n}(x)\!\!&=&\!\! \frac{\partial}{\partial x}\big\{h_{m}(x)g_n(x)\big\}
\nonumber\\
&=&\!\! \exp\left(-\frac{x^2}{2}\right)\sqrt{\frac{n!}{2^{m}m!}}
\,i^n \bigg\{(n+1) H_m(x)\left(D_{-2-n}\Big(i\sqrt{2}x\Big)
+(-1)^{n}D_{-2-n}\Big(-i\sqrt{2}x\Big)\right)\nonumber\\ &&\!\! 
+i\sqrt{2}m H_{m-1}(x)\left(D_{-1-n}\Big(i\sqrt{2}x\Big)-(-1)^{n}D_{-1-n}
\Big(-i\sqrt{2}x\Big)
\right)\bigg\}.
\end{eqnarray}
This form of the pattern functions is not symmetric with respect to
permutation of the indices $(m,n)$. It is identical with the pattern 
functions $F_{m,n}(x)$ in Eq.(4.25) only for $m \le n+1$ as discussed in
section 4 but can be taken as an equivalent pattern function for arbitrary
$(m,n)$ due to their nonuniqueness as discussed in section 6. Furthermore, 
it is clear that, instead of (7.16), one can take for $X_{k,l}^{(1)}(x)$ the  
combination with exchange of the normalizable and nonnormalizable
solutions of the wave equation, i.e.
\begin{equation}
X_{k,l}^{(1)}(x)= g_{k}(x)h_{l}(x), \quad Y_{m,n}(x)=h_{m}(x)h_{n}(x),
\end{equation}
or linear combinations of them. In case of (7.19) we find as the equivalent
pattern functions
\begin{equation}
F'''_{m,n}(x)=\frac{\partial}{\partial x}\big\{g_{m}(x)h_n(x)\big\} =
F''_{n,m}(x).
\end{equation}
These pattern functions are identical with the pattern
functions $F_{m,n}(x)$ considered in Eqs.(4.25) and (4.27) in the cases
\begin{equation}
F''_{m,n}(x)=F_{m,n}(x),\quad m \le n+1, \qquad F'''_{m,n}(x)=F_{m,n}(x),\quad
n\le m+1,
\end{equation}
and are connected with the pattern functions $F'_{m,n}(x)$ considered in
the last section by
\begin{equation}
F'_{m,n}(x)=\frac{1}{2}\left(F''_{m,n}(x)+F''_{n,m}(x)\right).
\end{equation}
Nevertheless, in all other cases of nonidentity one has completely 
equivalent pattern functions which can be used by same right as the 
pattern functions $F_{m,n}(x)$. As already mentioned, the pattern functions 
$F_{m,n}(x)$ possess the advantage that they vanish at infinity. In the 
special cases $m=n$ and $m=n\pm1$ we obtained the same pattern functions
by all considered methods. 

The nonuniqueness of the pattern functions may 
have as the cause a certain redundancy in the information contained in the 
Wigner quasiprobability and in its Radon transform but it seems to be 
difficult to eliminate this redundancy. The pattern functions themselves
play an auxiliary role and only the integrals (4.24) over the pattern 
functions leading to the Fock-state matrix elements possess an invariant
meaning. The representations of the pattern functions in the form (7.18) 
or (7.20) is useful for computer calculations. In comparison to (4.27) 
they possess the advantage that one has to calculate here only a sum over 
the real or imaginary parts of two functions of the parabolic cylinder 
of imaginary argument instead of maximally $n+1$ such functions there.

\setcounter{chapter}{8}
\setcounter{equation}{0}
\chapter*{8. Conclusion} 

We introduced in the present paper the Radon transform of the Wigner 
quasiprobability in its more general canonical representation in comparison 
to the usual and considered the reconstruction of the density operator via 
the Fock-state matrix elements and via the normally ordered moments. 
The inverse two-dimensional Radon transformation was considered 
in detail because some moments seem to have been clarified in the literature 
about quantum tomography. The transformation properties of the Wigner 
quasiprobability and its Radon and Fourier transforms with regard to 
displacement and squeezing of an initial state are considered with many 
technical details. This has a more practical aspect, for example, for the
calculation of the Radon transform of squeezed coherent states and 
squeezed-state excitations which becomes much easier by application of
the derived formulae. The calculation of the pattern functions for the 
reconstruction of the Fock-state matrix elements of the density operator 
via the normally ordered moments leads to a new representation of these 
functions by convergent series over Hermite polynomials of even or odd 
order which provides an alternative for calculation and plotting of these 
functions.

We did not consider in the present paper, for example, the influence of 
imperfect measurements which lead to some smoothing of the Radon transform 
with the problem to reconstruct the density operator from these smoothed 
Radon transforms, a problem, now intensively discussed in the literature. 
Furthermore, we did not extend the considerations to multi-mode cases but 
many features can be probably translated to these more general cases in a 
simple way. We hope that our considerations are useful for the clarification
of some principal problems connected with two-dimensional Radon transforms
and its application in quantum optics. \\[5mm]

\noindent{\large Acknowledgement}\\[3mm]
The author likes to express his gratitude for valuable discussions and hints
to 
V.P. Karassiov from Moscow, P.L. Knight from London, 
O.V. Man'ko and V.I. Man'ko from Moscow, T. Opatrn\'{y} from 
Olomouc, M.G.A. Paris from Pavia and Th. Richter from Berlin.


\begin{thebibliography}{99}
\bibitem{ulf} A problem posed by U. Leonhardt to me was to prove or disprove   
	      that the first maxima and minima of $F_{n,n}(x)$ 
	      are exactly equal to the value $\pm F_{n,n}(0)$. This can be 
	      disproved numerically by applying the expansion in Hermite 
	      polynomials (6.3). For example, $F_{2,2}(0)=2$, first
	      mimima $F_{2,2}(\pm 0.715045)= -2.00113$, first maxima
	      $F_{2,2}(\pm 1.54108)= 2.01617$.
\end{thebibliography}
\end{document}